\documentclass[aps,prl,reprint,showpacs,superscriptaddress,longbibliography]{revtex4-1}
\usepackage[dvipdfmx]{graphicx}

\usepackage{amsmath,amssymb,amsfonts}
\usepackage{comment}
\usepackage{color}
\usepackage[T1]{fontenc}
\usepackage{textcomp}
\usepackage{hyperref}
\usepackage{ulem}

\hypersetup{
setpagesize=false,
colorlinks=true,
linkcolor=blue,
citecolor=red,
}
\usepackage{bm}

\makeatletter
\let\MYcaption\@makecaption
\makeatother
\usepackage{subcaption}
\makeatletter
\let\@makecaption\MYcaption
\makeatother

\usepackage{physics}
\usepackage{comment}

\begin{document}
\title{%
Breaking down the magnonic Wiedemann-Franz law in the hydrodynamic regime}

\author{Ryotaro Sano}
\email{sano.ryotaro.52v@st.kyoto-u.ac.jp}
\affiliation{%
Department of Physics, Kyoto University, Kyoto 606-8502, Japan
}%

\author{Mamoru Matsuo}

\affiliation{%
Kavli Institute for Theoretical Sciences, University of Chinese Academy of Sciences, Beijing, 100190, China.
}%

\affiliation{%
CAS Center for Excellence in Topological Quantum Computation, University of Chinese Academy of Sciences, Beijing 100190, China
}%
\affiliation{%
RIKEN Center for Emergent Matter Science (CEMS), Wako, Saitama 351-0198, Japan
}%
\affiliation{%
Advanced Science Research Center, Japan Atomic Energy Agency, Tokai, 319-1195, Japan
}%

\date{\today}

\begin{abstract}
Recent experiments have shown an indication of a hydrodynamic magnon behavior in ultrapure ferromagnetic insulators; however, its direct observation is still lacking. Here, we derive a set of coupled hydrodynamic equations and study the thermal and spin conductivities for such a magnon fluid. We reveal the drastic breakdown of the magnonic Wiedemann-Franz law as a hallmark of the hydrodynamics regime, which will become key evidence for the experimental realization of an emergent hydrodynamic magnon behavior. Therefore, our results pave the way towards the direct observation of magnon fluids.
\end{abstract}

\maketitle

{\it Introduction.---}
Quantum transport has attracted a profound growth of interest owing to its fundamental importance and many applications. Recent significant developments in experimental techniques have further boosted the study of quantum transport. Notably in ultraclean systems, strong interactions between particles drastically affect the transport properties, resulting in an emergent hydrodynamic behavior: examples include electrons~\cite{Molenkamp1994,Amoretti2020PRR,Polini2020PhysToday}, phonons~\cite{GUO20151PhysRep,Cepellotti2015}, cold atoms~\cite{Cao2011Science,Filippone2016PRA,Dominik2018PNAS}, and quark-gluon plasmas~\cite{SHURYAK2004PPNP,Adamczyk2017}. In these systems, the conventional non-interacting description for particles breaks down where momentum-conserving scatterings become dominant, which in turn introduces a novel nonequilibrium state inherent in the so-called hydrodynamic regime. The most-studied example is the hydrodynamic charge transport in metals or semiconductors, which gives rise to an active research field called electron hydrodynamics~\cite{Moll2016Science,Bandurin2016,Levitov2016NatPhys,Alekseev2016PRL,KrishnaKumar2017NatPhys,Bandurin2018,Sulpizio2019,Gallagher2019,Ella2019NatNanotech,Jenkins2020arXiv,Vool2021NatPhys,Gusev2018PRB,Berdyugin2019Science,Aharon-Steinberg2022Nat,Guo2017PNAS,Keser2021PRX,Principi2015PRL,Crossno2016,Gooth2018,Jaoui2018NPJ,Zarenia2019PRB,Zarenia2019_2Dmater,Anderson2019PRL,ZareniaPRB2020}. This concept has revealed various unconventional transport phenomena such as the negative nonlocal resistance~\cite{Bandurin2016,Levitov2016NatPhys}, the Poiseuille flow~\cite{Ella2019NatNanotech,Jenkins2020arXiv,Vool2021NatPhys}, the Hall viscosity~\cite{Gusev2018PRB,Berdyugin2019Science,Aharon-Steinberg2022Nat}, the geometric control of the flow~\cite{Guo2017PNAS,Keser2021PRX}, and the violation of the Wiedemann-Franz (WF) law~\cite{Principi2015PRL,Crossno2016,Gooth2018,Jaoui2018NPJ,Zarenia2019PRB,Zarenia2019_2Dmater,Anderson2019PRL,ZareniaPRB2020}.

Recent experiments on ultrapure ferromagnetic insulators (FMI) have opened up new pathways for magnon hydrodynamics~\cite{vanderSar2015,Du2017Science,Prasai2017PRB}. Magnons in FMI attract special attention as a promising candidate for a spin information carrier~\cite{Kruglyak_2010,Serga_2010,Kajiwara2010,Chumak2015,Otani2017,Althammer_2018} with good coherence and without dissipation of the Joule heating compared to conduction electrons in metals~\cite{Uchida2010APL,Cornelissen2015Natphys,GoennenweinAPL2015,Li2016NatCommun,Demidov2017NatCommun,Olsson2020PRX,Schlitz2021PRL}. Therefore, hydrodynamic magnon transport implies exhibiting extraordinary features~\cite{Halperin1969PhysRev,Reiter1968PhysRev,MICHEL1969,Michel1970,Ulloa2019,Rodriguez2022PRB,Yuheng2022AIPAdvances} as well as electron hydrodynamics and has a potential for innovative functionalities beyond the conventional non-interacting magnon picture. However, the direct observation of magnon fluids remains an open issue due to the lack of probes to access the time and length scales characteristic of this regime.

In this Letter, we derive a set of coupled hydrodynamic equations for a magnon fluid in topologically trivial bulk FMI by focusing on the most dominant time scales. Based on the obtained equations, we investigate the thermal and spin conductivities for magnon systems in the hydrodynamic regime. In the conventional transport regime, the ratio between the two conductivities has a material-independent universal value, which is known as the magnonic Wiedemann-Franz (WF) law~\cite{Nakata2015PRB,Nakata2017PRB1,Nakata2017PRB2,Nakata2021,Nakata2022PRB}: a magnon-analog of the celebrated WF law~\cite{WF1853}. Here, as a hallmark of the hydrodynamic regime, we reveal that the ratio shows a large deviation from the law, implying that magnon-magnon interactions impact on the two conductivites in radically different ways. Therefore, our results are expected to become key evidence for the emergence of a hydrodynamic magnon behavior and lead to the direct observation of magnon fluids.

{\it Formulation.---}
We outline how to derive a set of coupled hydrodynamic equations for a magnon fluid (for details, see the Supplemental Material~\footnote{See the Supplemental Materials for the detailed derivation of the hydrodynamic equations, which includes Ref.~\cite{Haug2008}}). We start from the magnon Boltzmann equation~\cite{Zhang2004,Matsumoto2011PRB,Matsumoto2011PRL,Zhang2012PRL,zhang2012PRB,Rezende2014PRB,REZENDE2016jmagmag,Cornelissen2016,Schmidt2018,Liu2019PRB,Costa2019} which governs the evolution of the magnon distribution function $n_{\vb*{k}}(\vb*{r},t)$ in the phase space:
\begin{equation}
\pdv{n_{\vb*{k}}}{t}+\pdv{\omega_{\vb*{k}}}{\vb*{k}}\vdot\pdv{n_{\vb*{k}}}{\vb*{r}}
=\mathcal{C}^\mathrm{mm}_{\mathrm{N}}[n_{\vb*{k}}]-\frac{n_{\vb*{k}}-N_{\vb*{k}}}{\tau_\mathrm{U}},\label{Boltzmann}
\end{equation}
where $\omega_{\vb*{k}}$ is the dispersion of magnons with momentum $\vb*{k}$. For ultrapure FMI, we only consider magnon-number conserving exchange interactions as magnon-magnon scattering processes, which can be divided into the contribution of normal (N) process and Umklapp (U) one in the usual manner, and neglect the dipolar interactions. We have also assumed the absence of external driving forces and relaxation time approximation for U process with $\tau_\mathrm{U}$. $\mathcal{C}_\mathrm{N}^\mathrm{mm}$ is the collision integral for N process. Here, N process conserves the momentum, while U process does not. As N scattering rates are very large, an out-of-equilibrium distribution $n_{\vb*{k}}$ will decay first into the drifting distribution $n_{\vb*{k}}^{(0)}$ given by Eq.~\eqref{local:eq}, and from this state it will relax towards static thermal equilibrium. Instead, U process tends to relax the magnon populations towards local thermal equilibrium $N_{\vb*{k}}$ which is given by the Bose-Einstein distribution with a finite chemical potential $\mu$ due to the magnon number conservation.

From the collision integral $\mathcal{C}^\mathrm{mm}_{\mathrm{N}}$ in Eq.~\eqref{Boltzmann}, we can identify the collision invariants $\mathcal{N}_\lambda$, which are defined as $\int[\dd{\vb*{k}}]\mathcal{C}^\mathrm{mm}_\mathrm{N}\mathcal{N}_\lambda=0$ with $\int[\dd{\vb*{k}}]\equiv\int\dd[3]{\vb*{k}}/(2\pi)^3$. $\mathcal{C}^\mathrm{mm}_\mathrm{N}$ guarantees that the quantities $\mathcal{N}_\lambda=(1,\hbar\vb*{k},\hbar\omega_{\vb*{k}})$ do not change in the evolution of the distribution function. Following the standard approach~\cite{Landau_Kinetics,Lucas_2018,NAROZHNY2019167979,Narozhny2019,Narozhny2021,Kiselev2020,Narozhny2022}, the conservation laws for number, momentum, and energy densities are obtained as follows:
\begin{subequations}
\begin{align}
    \pdv{t}\langle\mathcal{N}_0\rangle&+\div\vb*{\mathcal{J}}^{\mathcal{N}_0}=0,\\
    \pdv{t}\langle\mathcal{N}_i\rangle&+\div\vb*{\mathcal{J}}^{\mathcal{N}_i}=-\frac{\langle\mathcal{N}_i\rangle}{\tau_{\mathrm{U}}},\\
    \pdv{t}\langle\mathcal{N}_4\rangle&+\div\vb*{\mathcal{J}}^{\mathcal{N}_4}=0,
\end{align}
\end{subequations}
where we have defined the fluxes $\vb*{\mathcal{J}}^{\mathcal{N}_\lambda}$ corresponding to each invariant density $\langle\mathcal{N}_\lambda\rangle\equiv\int[\dd{\vb*{k}}]\mathcal{N}_\lambda n_{\vb*{k}}$. $-\langle\mathcal{N}_i\rangle/\tau_\mathrm{U}$ is the momentum relaxation force stems from U process.

In the hydrodynamic regime, the system reaches local equilibrium via N magnon-magnon scatterings, which conserve the total number, momentum, and energy of the magnon system. For this reason, we assume that the distribution functions in the zeroth order are described as
\begin{equation}
    n_{\vb*{k}}^{(0)}=\left[\exp\left\{\frac{\hbar\omega_{\vb*{k}}-\hbar\vb*{k}\vdot\vb*{v}-\tilde{\mu}}{k_\mathrm{B}T}\right\}-1\right]^{-1},\label{local:eq}
\end{equation}
which is referred to as the local equilibrium distribution function. Notably, $(-\tilde{\mu}/T,-\vb*{v}/T,1/T)$ are the position- and time-dependent intensive thermodynamic parameters, which are conjugate to each collision invariant $\mathcal{N}_\lambda=(1,\hbar\vb*{k},\hbar\omega_{\vb*{k}})$ of the magnon system. Here, the drift velocity $\vb*{v}$ is a Lagrange multiplier enforcing the momentum conservation. $\tilde{\mu}$ is the chemical potential in the frame which moves with the fluid: $\tilde{\mu}=\mu-m^\ast\vb*{v}^2/2$. We use the quadratic dispersion for magnons $\hbar\omega_{\vb*{k}}=\hbar^2\vb*{k}^2/2m^\ast$ with the effective mass $m^\ast$ for similicity and conputability~\cite{Rodriguez2022PRB}. Under these conditions, $n^{(0)}_{\vb*{k}}$ is transformed into $N_{\vb*{k}}$ under the Galilean transformation from the frame in which the fluid moves with the velocity $\vb*{v}$ to the frame in which it is at rest: $\hbar\vb*{k}\to\hbar\vb*{k}+m^\ast\vb*{v}$. Note that the non-equilibrium magnon chemical potential $\mu$~\cite{Cornelissen2016,Du2017Science,Demidov2017NatCommun,Olsson2020PRX,Schlitz2021PRL} is present here due to the magnon number conservation under both N and U processes. The magnon chemical potential is an essential ingredient for heat and spin transport.

Deviation from the local equilibrium distribution $n_{\vb*{k}}'=n_{\vb*{k}}-n_{\vb*{k}}^{(0)}$ creates the dissipative number, momentum, and energy fluxes: $\vb*{J}=\vb*{J}^{(0)}+\vb*{J}'$, $\Pi_{ji}=\Pi_{ji}^{(0)}+\Pi'_{ji}$, and $\vb*{Q}=\vb*{Q}^{(0)}+\vb*{Q}'$ with $\vb*{J}^{(0)}=0$, $\Pi_{ji}^{(0)}=P\delta_{ji}$, and $\vb*{Q}^{(0)}=0$. By using these quantities, the hydrodynamic equations are obtained as follows:
\begin{subequations}\label{hydro:Eq}
\begin{gather}
\pdv{n}{t}+\div(n\vb*{v}+\vb*{J}')=0,\label{number}\\
\rho\left(\pdv{t}+\vb*{v}\vdot\grad\right)v_i+\pdv{x_i}P+\pdv{x_j}\Pi_{ji}'=-\frac{\rho{v}_i}{\tau_\mathrm{U}},\label{momentum}\\
\pdv{u}{t}+\div(u\vb*{v}+\vb*{Q}')+P\div\vb*{v}+\Pi'_{ji}\pdv{v_i}{x_j}=\frac{\rho\vb*{v}^2}{\tau_\mathrm{U}},\label{energy}
\end{gather}
\end{subequations}
where $n$, $\rho\vb*{v}=m^\ast n\vb*{v}$, and $u$ are the magnon number, momentum, and internal energy densities. The force that drives the magnon flow is the gradient of the magnon pressure $P$. Heat currents $\vb*{Q}=\vb*{Q}'$ can be obtained only when considering the deviation from the local equilibrium. Here, U process produces not only the momentum relaxation but also the internal energy relaxation to the lattice environment. Although similar hydrodynamic equations are derived in Ref.~\cite{Rodriguez2022PRB} from the same starting point Eq.~\eqref{local:eq}, the inequivalence of the momentum density and the number flux stems from fluctuations around the local equilibrium here, which is different from the correction to the particle current operator due to the magnon-magnon interactions~\cite{Rodriguez2022PRB}. Furthermore, our theory includes the momentum relaxation processes which have a significant importance for the main results.

{\it Entropy production.---}
The entropy production is one of intriguing aspects of the hydrodynamic approach~\cite{chaikin_lubensky_1995}. After straightforward calculations, the entropy production rate is obtained as the balance equation for the entropy density $s=u+P-\mu n$:
\begin{align}
    &\pdv{s}{t}+\div\left(s\vb*{v}+\frac{\vb*{Q}'-\mu\vb*{J}'}{T}\right)\nonumber\\
    &=\vb*{Q}'\vdot\grad\left(\frac{1}{T}\right)+\frac{\Pi_{ji}'}{T}\partial_j(-v_i)+\vb*{J}'\vdot\grad\left(-\frac{\mu}{T}\right)\nonumber\\
    &\quad+\frac{1}{T}\frac{\rho\vb*{v}^2}{\tau_\mathrm{U}}.\label{entropy}
\end{align}
The entropy production rate is given by the product of thermodynamic driving forces and their conjugate dissipative currents. We should note that the adiabatic evolution condition, $\vb*{Q}'=\vb*{J}'=\Pi_{ji}'=0$, reproduces the dissipationless equations in Eqs.~\eqref{hydro:Eq}. Furthermore, these pairs obey the Onsager reciprocity relations~\cite{Onsager1931PhysRev1,Onsager1931PhysRev2}. These relations describe the effects on an irreversible flux of an extensive conserved quantity by intensive thermodynamic variables. For example, gradients in the magnon chemical potential can be treated as conjugate forces for spin transport in the Onsager relations. By analogy with the thermo-electric Onsager relations, thermo-spin Onsager relations connect magnonic spin and heat currents.

{\it Thermal and spin conductivities.---}
Based on the obtained hydrodynamic equations Eqs.~\eqref{hydro:Eq} and the entropy production rate Eq. ~\eqref{entropy}, we investigate the spin and heat currents in the hydrodynamic regime.
In the following analysis, we consider the linear response under constant gradients of the temperature and chemical potential. In order to calculate the dissipative heat and magnon currents, we also perform the relaxation time approximation for N process,
\begin{equation}
    \mathcal{C}^\mathrm{mm}_\mathrm{N}[n_{\vb*{k}}]=-\frac{n_{\vb*{k}}-n^{(0)}_{\vb*{k}}}{\tau_\mathrm{N}}.
\end{equation}
For the purpose of obtaining the linear thermal conductivity, we assume the deviation from the local equilibrium distribution $\delta n_{\vb*{k}}=n_{\vb*{k}}-n_{\vb*{k}}^{(0)}$ is small to justify the use of the linearized Boltzmann equation. Then, $\delta n_{\vb*{k}}$ is obtained as
\begin{align}
    \delta n_{\vb*{k}}&\simeq-\left(\frac{1}{\tau_\mathrm{N}}+\frac{1}{\tau_\mathrm{U}}\right)^{-1}\nonumber\\
    &\times\left[\left(\pdv{t}+\pdv{\omega_{\vb*{k}}}{\vb*{k}}\vdot\grad\right)n_{\vb*{k}}^{(0)}+\frac{n_{\vb*{k}}^{(0)}-N_{\vb*{k}}}{\tau_\mathrm{U}}\right].
\end{align}
We further assume that $\delta n_{\vb*{k}}$ satisfies the conditions: $\int[\dd{\vb*{k}}]\mathcal{N}_\lambda\delta n_{\vb*{k}}=0$. Under these conditions, the densities of conserved quantities $\langle\mathcal{N}_\lambda(\vb*{r},t)\rangle=\langle\mathcal{N}_\lambda(\vb*{r},t)\rangle^{(0)}$ are determined by the local equilibrium distribution $n_{\vb*{k}}^{(0)}$. This assumption is necessary for not violating the conservation laws under the relaxation time approximation~\cite{Hattori2022symm}.

In the following analysis, we ignore $\partial/\partial t$ because we are interested in steady state transport properties. Substituting $\delta n_{\vb*{k}}$ into hydrodynamic variables, we obtain the magnon and heat currents in the linear response regime respectively:
\begin{subequations}
\begin{align}
    \vb*{J}&=-\frac{\tau_\mathrm{N}}{\tau_\mathrm{N}+\tau_\mathrm{U}}n_\mathrm{m}^{(0)}\vb*{v}+\left(\frac{1}{\tau_\mathrm{N}}+\frac{1}{\tau_\mathrm{U}}\right)^{-1}\frac{T}{m^\ast}\nonumber\\
    &\times\biggl[n_\mathrm{m}^{(0)}\grad\left(-\frac{\mu}{T}\right)+(u^{(0)}+P)\grad\left(\frac{1}{T}\right)\biggr],\\
    \vb*{Q}&=-\frac{\tau_\mathrm{N}}{\tau_\mathrm{N}+\tau_\mathrm{U}}(u^{(0)}+P)\vb*{v}+\left(\frac{1}{\tau_\mathrm{N}}+\frac{1}{\tau_\mathrm{U}}\right)^{-1}\frac{T}{m^\ast}\nonumber\\
    &\times\biggl[(u^{(0)}+P)\grad\left(-\frac{{\mu}}{T}\right)+\frac{7}{3}\langle\mathcal{N}_4^2\rangle_\mathrm{eq}\grad\left(\frac{1}{T}\right)\biggr],
\end{align}
\end{subequations}
where $\langle\mathcal{N}_4^2\rangle_\mathrm{eq}\equiv\int[\dd{\vb*{k}}](\hbar\omega_{\vb*{k}})^2N_{\vb*{k}}$. The transport coefficients $\mathcal{L}_{ij}$ in the linear response regime are defined as,
\begin{equation}
    \left(
    \begin{array}{c}
        \vb*{J}   \\
        \vb*{Q}
    \end{array}
    \right)=\left(
    \begin{array}{cc}
        \mathcal{L}_{11} & \mathcal{L}_{12} \\
        \mathcal{L}_{21} & \mathcal{L}_{22}
    \end{array}
    \right)\left(
    \begin{array}{c}
         \grad(-\mu/T)  \\
         \grad(1/T)
    \end{array}
    \right).\nonumber
\end{equation}
In the hydrodynamic regime, the transport coefficients are described by the drift velocity $\vb*{v}$. In our hydrodynamic equations, $\grad P=-n_\mathrm{m}^{(0)}T[\grad(-\mu/T)+\alpha\grad(1/T)]$ drives magnon flows with $\alpha=(u^{(0)}+P)/n_\mathrm{m}^{(0)}$. By solving Eq.~\eqref{momentum} as $\rho_\mathrm{m}^{(0)}\vb*{v}=\tau_\mathrm{U}\mathcal{V}n_\mathrm{m}^{(0)}T[\grad(-\mu/T)+\alpha\grad(1/T)]$, the components of $\mathcal{L}_{ij}$ are obtained as,
\begin{subequations}\label{coeff}
    \begin{align}
        \mathcal{L}_{11}&=\left(\frac{1}{\tau_\mathrm{N}}+\frac{1}{\tau_\mathrm{U}}\right)^{-1}n_\mathrm{m}^{(0)}\frac{T}{m^\ast}\left(1-\mathcal{V}\right),\\
        \mathcal{L}_{12}&=\left(\frac{1}{\tau_\mathrm{N}}+\frac{1}{\tau_\mathrm{U}}\right)^{-1}(u^{(0)}+P)\frac{T}{m^\ast}\left(1-\mathcal{V}\right)\\
        &=\mathcal{L}_{21}=\alpha\mathcal{L}_{11},\\
        \mathcal{L}_{22}&=\left(\frac{1}{\tau_\mathrm{N}}+\frac{1}{\tau_\mathrm{U}}\right)^{-1}(u^{(0)}+P)\frac{T}{m^\ast}\nonumber\\
        &\quad\times\left(\frac{7}{2}k_\mathrm{B}T\frac{\mathrm{Li}_{7/2}(z)}{\mathrm{Li}_{5/2}(z)}-\mathcal{V}\alpha\right),
    \end{align}
\end{subequations}
where we have introduced the polylogarithm function $\mathrm{Li}_s(z)\equiv\sum_{n=1}^\infty z^n/n^s$, the fugacity $z\equiv e^{\mu/k_\mathrm{B}T}$, and the renormalized velocity $\mathcal{V}$. $\mathcal{V}$ deviates from unity when viscous effects are switched on. As can be seen from Eqs.~\eqref{coeff}, the transport coefficients $\mathcal{L}_{ij}$ obey the Onsager reciprocity relations.

In the following, we consider magnon fluids in bulk FMI in order to justify neglecting nonlocal effects. Substituting $\mathcal{V}\approx1$ in Eqs.~\eqref{coeff}, we obtain the spin and thermal conductivities as
\begin{subequations}\label{conductivities}
\begin{align}
    \sigma_\mathrm{m}&=\tau_\mathrm{U}\frac{1}{m^\ast}\frac{\hbar}{\Lambda_{T}^3}\mathrm{Li}_{3/2}(z),\\
    \kappa_\mathrm{m}&=\left(\frac{1}{\tau_\mathrm{N}}+\frac{1}{\tau_\mathrm{U}}\right)^{-1}\frac{1}{m^\ast}\frac{k_\mathrm{B}^2T}{\Lambda_{T}^3}\left[\frac{35}{4}\mathrm{Li}_{7/2}(z)-\frac{25}{4}\frac{\mathrm{Li}_{5/2}^2(z)}{\mathrm{Li}_{3/2}(z)}\right],
\end{align}
\end{subequations}
where $\Lambda_{T}\equiv\sqrt{2\pi\hbar/m^\ast k_\mathrm{B}T}$ is the thermal de Broglie wavelength. Here, we have defined the spin and thermal conductivities as $\hbar(n_\mathrm{m}^{(0)}\vb*{v}+\vb*{J})=\sigma_\mathrm{m}\grad(-{\mu})$ and $\vb*{Q}=\kappa_\mathrm{m}\grad(-T)$. While the spin conductivity is not affected by momentum-conserving N process because the spin current here is the momentum flow of magnons, the thermal conductivity is drastically changed by both processes as a hallmark of the hydrodynamic regime.

{\it Breakdown of the magnonic WF law.---}
We are now ready to discuss the breakdown of the magnonic WF law in the hydrodynamic regime. The magnonic WF law claims that the thermal and spin conductivities satisfy the relation~\cite{Nakata2015PRB,Nakata2017PRB1,Nakata2017PRB2,Nakata2021,Nakata2022PRB},
\begin{align}
\left.\frac{\kappa_\mathrm{m}}{\sigma_\mathrm{m}T}\right|_{\mathrm{non-int}}&=\frac{k_\mathrm{B}^2}{\hbar}\left[\frac{35}{4}\frac{\zeta(7/2)}{\zeta(3/2)}-\frac{25}{4}\frac{\zeta^2(5/2)}{\zeta^2(3/2)}\right],
\end{align}
where this ratio shows the material-independent universal value so-called the magnonic Lorenz number: a magnon-analog of the well-known WF law~\cite{WF1853}. Here, we have used the fact that the polylogarithm function $\mathrm{Li}_s(z)$ can be approximated by the Riemann zeta function $\zeta(s)=\sum_{n=1}^\infty1/n^s$ for $z\approx1$. In the hydrodynamic regime, the ratio becomes
\begin{equation}
    \left.\frac{\kappa_\mathrm{m}}{\sigma_\mathrm{m}T}\right|_\mathrm{hydro}=\left(\frac{\tau_\mathrm{N}}{\tau_\mathrm{N}+\tau_\mathrm{U}}\right)\left.\frac{\kappa_\mathrm{m}}{\sigma_\mathrm{m}T}\right|_{\mathrm{non-int}},\label{hydroWF}
\end{equation}
which indicates that the standard form of the magnonic WF law is reduced by a factor $\tau_\mathrm{N}/(\tau_\mathrm{N}+\tau_\mathrm{U})$ owing to the difference in the relaxation processes between spin and heat currents in Eq.~\eqref{conductivities}. Eq.~\eqref{hydroWF} is the main result of this Letter. Especially in the hydrodynamic regime, $\tau_\mathrm{N}\ll\tau_\mathrm{U}$, large deviations from the law are implied by this factor.

\begin{figure}[t]
    \centering
    \includegraphics[keepaspectratio,scale=0.35]{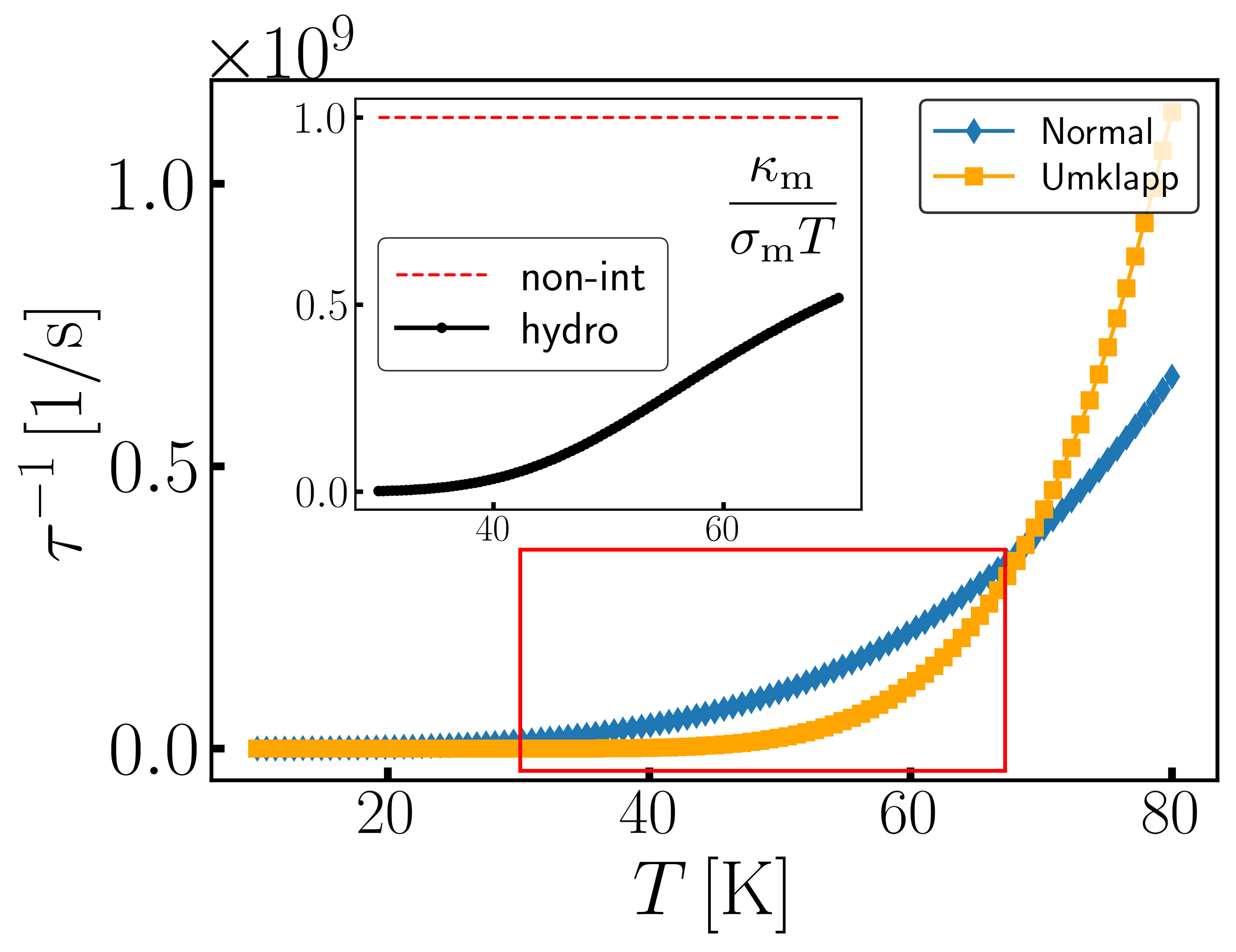}
    \caption{Breaking down the magnonic Wiedemann-Franz law. Temperature dependence of relaxation rates $\tau^{-1}$ for both Normal and Umklapp magnon-magnon scatterings are depicted. The red box shows the area satisfying $\tau_\mathrm{N}\ll\tau_\mathrm{U}$, namely, the hydrodynamic regime. The inset shows the ratio $\kappa_\mathrm{m}/\sigma_\mathrm{m}T|_\mathrm{hydro}$, calculated from Eq.~\eqref{hydroWF} in units of the magnonic Lorenz number $\kappa_\mathrm{m}/\sigma_\mathrm{m}T|_\mathrm{non-int}$. When the system enters the hydrodynamic regime, the ratio shows a large deviation from unity.}
    \label{fig:WF}
\end{figure}

{\it Discussion.---}
Finally, we will discuss the experimental feasibility of the breakdown of the magnonic WF law. According to Refs.~\cite{Bhandari1966PhysRev,Boona2014PRB,Bender2014PRB,Ruckriege2014PRB}, relaxation times are estimated as $\tau_\mathrm{N}^{-1}\sim(T/T_\mathrm{e})^3k_\mathrm{B}T/\hbar$ and $\tau_\mathrm{U}^{-1}\sim\sqrt{T/T_\mathrm{e}}\exp(-12T_\mathrm{e}/T)k_\mathrm{B}T/\hbar$ with $T_\mathrm{e}=2SJ/k_\mathrm{B}\sim37$\,K for yttrium-iron-garnet (YIG). The difference in temperature dependences of these two relaxation times originates from possible regions in reciprocal space under the scattering processes limited by overall conservation of energy and momentum modulo reciprocal lattice vectors~\cite{MICHEL1969,Michel1970,Ziman1960,Forney1973}. Fig.~[\ref{fig:WF}] illustrates the temperature dependences of scattering rates $1/\tau$. The red box dictates the area satisfying $\tau_\mathrm{N}\ll\tau_\mathrm{U}$, namely, the hydrodynamic regime. The ratio $\kappa_\mathrm{m}/\sigma_\mathrm{m}T|_\mathrm{hydro}$ in units of the magnonic Lorenz number $\kappa_\mathrm{m}/\sigma_\mathrm{m}T|_\mathrm{non-int}$ is depicted in the inset of Fig.~[\ref{fig:WF}]. When the magnon system enters the hydrodynamic regime, the ratio shows a large deviation from unity, which in turn results in the drastic breakdown of the magnonic WF law. We should discuss the effect of the dipolar interactions in realistic magnets which violates the magnon number conservation. The magnon dumping rate due to the dipolar interactions is in the order of $10^{6}$\,s$^{-1}$~\cite{Rodriguez2022PRB} and is much slower than $\tau^{-1}_\mathrm{N}$ and $\tau_{\mathrm{U}}^{-1}$ [see Fig.~\ref{fig:WF}]. Therefore, we neglect the dipolar interactions here and the hydrodynamic description which includes the number conservation law is valid under the dynamics we are interested in.

As the system enters the collisionless regime where the hydrodynamic description is invalid, an emergent sound mode referred to as the zero-magnon mode appears in low dimensional Heisenberg ferromagnets~\cite{Goll2020PRB}. Although it needs microscopic calculations and is beyond the scope of our study, the impact of the zero-magnon mode on the magnonic Lorenz number may be an interesting future work.

{\it Conclusion.---}%
In summary, we have developed a basic framework of magnon hydrodynamics for topologically trivial bulk FMI, which is composed of the hydrodynamic equations Eqs.~\eqref{hydro:Eq} and the entropy production rate Eq.~\eqref{entropy}. Based on these equations, we have first investigated the dissipative magnon and heat currents, which obey the Onsager reciprocity relations, and then the thermal and spin conductivities. As a hallmark of the hydrodynamic regime, we have revealed that the ratio between the two conductivities shows a large deviation from the standard form so-called the magnonic WF law. Here, we have identified an origin of the drastic breakdown as the difference in relaxation processes between spin and heat currents, which is unique to the hydrodynamic regime. The violation of the magnonic WF law does not depend on the details of the magnon dispersion: a universal feature of the hydrodynamic regime. Therefore, our results may yield key evidence for the experimental realization and detection of magnon fluids in a wide range of materials.

\begin{acknowledgements}
The authors are grateful to D. Oue, K. Shinada, T. Funato, J. Fujimoto, and Y. Suzuki for valuable discussions. We also thank Y. Nozaki, K. Yamanoi, and T. Horaguchi for providing helpful comments from an experimental point of view. The authors also thank the referee for noticing Ref.~\cite{Goll2020PRB}. R.S. is supported by JSPS KAKENHI (Grants JP 22J20221). This work was supported by the Priority Program of Chinese Academy of Sciences under Grant No. XDB28000000, and by JSPS KAKENHI for Grants (Nos. 20H01863 and 21H04565) from MEXT, Japan.
\end{acknowledgements}

\bibliography{ref}

\clearpage

\title{Supplemental Materials for \\``Breaking down the magnonic Wiedemann-Franz law in hydrodynamic regime''}

\author{Ryotaro Sano}
\affiliation{%
	Department of Physics, Kyoto University, Kyoto 606-8502, Japan
}%

\author{Mamoru Matsuo}
%\email{}
\affiliation{%
	Kavli Institute for Theoretical Sciences, University of Chinese Academy of Sciences, Beijing, 100190, China.
}%

\affiliation{%
	CAS Center for Excellence in Topological Quantum Computation, University of Chinese Academy of Sciences, Beijing 100190, China
}%
\affiliation{%
	RIKEN Center for Emergent Matter Science (CEMS), Wako, Saitama 351-0198, Japan
}%
\affiliation{%
	Advanced Science Research Center, Japan Atomic Energy Agency, Tokai, 319-1195, Japan
}%

\maketitle
\renewcommand{\theequation}{S.\arabic{equation}}
\onecolumngrid

\newcommand{\D}[2][{}]{\frac{D^{#1}{#2}}{Dt^{#1}}}
\newcommand{\Du}[2][{}]{\frac{D_{\vb*{u}}^{#1}{#2}}{Dt^{#1}}}
\newcommand{\Dv}[2][{}]{\frac{D_{\vb*{v}}^{#1}{#2}}{Dt^{#1}}}

\section{Derivation of the hydrodynamic equations for Magnon Fluids}
\subsection{I. Boltzmann Equation}
In this section, we outline how to derive the conservation laws for magnon fluids. We start from the magnon Boltzmann equation,
\begin{equation}
	\pdv{n_{\vb*{k}}(\vb*{r},t)}{t}+\dot{\vb*{r}}\vdot\pdv{n_{\vb*{k}}(\vb*{r},t)}{\vb*{r}}+\dot{\vb*{k}}\vdot\pdv{n_{\vb*{k}}(\vb*{r},t)}{\vb*{k}}=\mathcal{C}[n_{\vb*{k}}],\label{S:Boltzmann}
\end{equation}
which governs the evolution of the magnon distribution function $n_{\vb*{k}}(\vb*{r},t)$ in the phase space. The semiclassical equations of motion for magnons read
\begin{equation}
	\dot{\vb*{r}}=\pdv{\omega_{\vb*{k}}}{\vb*{k}},\qquad\dot{\vb*{k}}=0,
\end{equation}
where the external driving force is absent for simplicity. $\hbar\omega_{\vb*{k}}$ is the energy of the magnon with the momentum $\vb*{k}$. The collision integral $\mathcal{C}$ is divided into the contribution of normal (N) process and Umklapp (U) one in the usual manner,
\begin{equation}
	\mathcal{C}[n_{\vb*{k}}]=\mathcal{C}^\mathrm{mm}_\mathrm{N}[n_{\vb*{k}}]+\mathcal{C}^\mathrm{mm}_\mathrm{U}[n_{\vb*{k}}].
\end{equation}
Note that we only consider magnon number-conserving scatterings in the following. For an interacting magnon gas, the collision integral for N process under the first Born approximation is given by ~\cite{Haug2008}
\begin{equation}
	\mathcal{C}^\mathrm{mm}_\mathrm{N}[n_{\vb*{k}}]=\int[\dd{\vb*{k}'}]\int[\dd{\vb*{k}_1}]\int[\dd{\vb*{k}_1'}]w^\mathrm{mm}_\mathrm{N}(\vb*{k},\vb*{k}_1;\vb*{k}',\vb*{k}'_1)[(1+n_{\vb*{k}})(1+n_{\vb*{k}_1})n_{\vb*{k}'}n_{\vb*{k}_1'}-n_{\vb*{k}}n_{\vb*{k}_1}(1+n_{\vb*{k}'})(1+n_{\vb*{k}_1'})],
\end{equation}
where the intrinsic transition probability per unit time is
\begin{equation}
	w^\mathrm{mm}_\mathrm{N}(\vb*{k},\vb*{k}_1;\vb*{k}',\vb*{k}'_1)=\frac{2\pi}{\hbar}\frac{1}{2}\abs{W^\mathrm{mm}(\vb*{k},\vb*{k}_1;\vb*{k}',\vb*{k}'_1)+W^\mathrm{mm}(\vb*{k}_1,\vb*{k},\vb*{k}_1',\vb*{k}')}^2\delta(\hbar\omega_{\vb*{k}}+\hbar\omega_{\vb*{k}_1}-\hbar\omega_{\vb*{k}'}-\hbar\omega_{\vb*{k}'_1})\delta(\vb*{k}+\vb*{k}_1-\vb*{k}'-\vb*{k}'_1).\label{CI_N:S}
\end{equation}
Here, $W(\vb*{k},\vb*{k}_1;\vb*{k}',\vb*{k}_1')=\matrixel{\vb*{k},\vb*{k}_1}{\hat{W}}{\vb*{k}',\vb*{k}_1'}$ is the interaction matrix element. Eq.~\eqref{CI_N:S} shows that N process conserves both energy and momentum during the collisions. On the other hand, we assume the relaxation approximation for the U process which does not conserve the momentum,
\begin{equation}
	\mathcal{C}^\mathrm{mm}_\mathrm{U}[n_{\vb*{k}}]=-\frac{n_{\vb*{k}}(\vb*{r},t)-N_{\vb*{k}}(\vb*{r},t)}{\tau_\mathrm{U}}.
\end{equation}
Therefore, U processes tend to relax the magnon populations towards local thermal equilibrium $N_{\vb*{k}}$.

We introduce an arbitrary function $\mathcal{N}(\vb*{k},n_{\vb*{k}})$ which depends on the momentum $\vb*{k}$ and the distribution $n_{\vb*{k}}$. Its local density is defined as
\begin{equation}
	\langle\mathcal{N}(\vb*{r},t)\rangle\equiv\int[\dd{\vb*{k}}]\mathcal{N}(\vb*{k},n_{\vb*{k}})n_{\vb*{k}}.
\end{equation}
The change of this function due to the N collisions is calculated as
\begin{align}
	\left.\pdv{\langle\mathcal{N}(\vb*{r},t)\rangle}{t}\right|_{\mathrm{coll:N}}&=\int[\dd{\vb*{k}}]\left.\pdv{[\mathcal{N}(\vb*{k},n_{\vb*{k}})n_{\vb*{k}}]}{t}\right|_{\mathrm{coll:N}}=\int[\dd{\vb*{k}}]\mathcal{C}^\mathrm{mm}_\mathrm{N}[n_{\vb*{k}}]\pdv{[\mathcal{N}(\vb*{k},n_{\vb*{k}})n_{\vb*{k}}]}{n_{\vb*{k}}}\nonumber\\
	&=\int[\dd{\vb*{k}}]\int[\dd{\vb*{k}'}]\int[\dd{\vb*{k}_1}]\int[\dd{\vb*{k}_1'}]w_\mathrm{N}^\mathrm{mm}(\vb*{k},\vb*{k}_1;\vb*{k}',\vb*{k}_1')\pdv{[\mathcal{N}(\vb*{k},n_{\vb*{k}})n_{\vb*{k}}]}{n_{\vb*{k}}}\nonumber\\
	&\times[(1+n_{\vb*{k}})(1+n_{\vb*{k}_1})n_{\vb*{k}'}n_{\vb*{k}_1'}-n_{\vb*{k}}n_{\vb*{k}_1}(1+n_{\vb*{k}'})(1+n_{\vb*{k}_1'})].\label{S:Boltzmann1}
\end{align}
Exploiting the symmetry of the intrinsic transition probability $w_\mathrm{N}^\mathrm{mm}(\vb*{k},\vb*{k}_1;\vb*{k}',\vb*{k}_1')$ with respect to the exchange of particle coordinates
\begin{subequations}
	\begin{align}
		w_\mathrm{N}^\mathrm{mm}(\vb*{k},\vb*{k}_1;\vb*{k}',\vb*{k}_1')&=w_\mathrm{N}^\mathrm{mm}(\vb*{k}_1,\vb*{k};\vb*{k}'_1,\vb*{k}')\\
		&=w_\mathrm{N}^\mathrm{mm}(\vb*{k}',\vb*{k}'_1;\vb*{k},\vb*{k}_1)=w_\mathrm{N}^\mathrm{mm}(\vb*{k}_1',\vb*{k}';\vb*{k}_1,\vb*{k}),
	\end{align}\label{symmetry}
\end{subequations}
Eq.~\eqref{S:Boltzmann1} is rewritten as
\begin{align}
	\int[\dd{\vb*{k}}]\mathcal{C}^\mathrm{mm}_\mathrm{N}[n_{\vb*{k}}]\pdv{[\mathcal{N}(\vb*{k},n_{\vb*{k}})n_{\vb*{k}}]}{n_{\vb*{k}}}&=\frac{1}{4}\int[\dd{\vb*{k}}]\int[\dd{\vb*{k}'}]\int[\dd{\vb*{k}_1}]\int[\dd{\vb*{k}_1'}]w_\mathrm{N}^\mathrm{mm}(\vb*{k},\vb*{k}_1;\vb*{k}',\vb*{k}_1')\nonumber\\
	&\times\left\{\pdv{[\mathcal{N}(\vb*{k},n_{\vb*{k}})n_{\vb*{k}}]}{n_{\vb*{k}}}+\pdv{[\mathcal{N}(\vb*{k}_1,n_{\vb*{k}_1})n_{\vb*{k}_1}]}{n_{\vb*{k}_1}}-\pdv{[\mathcal{N}(\vb*{k}',n_{\vb*{k}'})n_{\vb*{k}'}]}{n_{\vb*{k}'}}-\pdv{[\mathcal{N}(\vb*{k}'_1,n_{\vb*{k}_1'})n_{\vb*{k}_1'}]}{n_{\vb*{k}_1'}}\right\}\nonumber\\
	&\times[(1+n_{\vb*{k}})(1+n_{\vb*{k}_1})n_{\vb*{k}'}n_{\vb*{k}_1'}-n_{\vb*{k}}n_{\vb*{k}_1}(1+n_{\vb*{k}'})(1+n_{\vb*{k}_1'})].\label{S:Boltzmann2}
\end{align}

\subsection{II. H-Theorem}
Now we consider the following choice for $\mathcal{N}$:
\begin{equation}
	\mathcal{N}(\vb*{k},n_{\vb*{k}})n_{\vb*{k}}=(1+n_{\vb*{k}})\log(1+n_{\vb*{k}})-n_{\vb*{k}}\log n_{\vb*{k}}.
\end{equation}
The partial derivative with respect to $n_{\vb*{k}}$ yields
\begin{equation}
	\pdv{[\mathcal{N}(\vb*{k},n_{\vb*{k}})n_{\vb*{k}}]}{n_{\vb*{k}}}=\log\frac{n_{\vb*{k}}}{1+n_{\vb*{k}}},
\end{equation}
and Eq.~\eqref{S:Boltzmann2} becomes
\begin{align}
	\int[\dd{\vb*{k}}]\mathcal{C}^\mathrm{mm}_\mathrm{N}[n_{\vb*{k}}]\pdv{[\mathcal{N}(\vb*{k},n_{\vb*{k}})n_{\vb*{k}}]}{n_{\vb*{k}}}&=-\frac{1}{4}\int[\dd{\vb*{k}}]\int[\dd{\vb*{k}'}]\int[\dd{\vb*{k}_1}]\int[\dd{\vb*{k}_1'}]w_\mathrm{N}^\mathrm{mm}(\vb*{k},\vb*{k}_1;\vb*{k}',\vb*{k}_1')\log\left[\frac{(1+n_{\vb*{k}})(1+n_{\vb*{k}_1})n_{\vb*{k}'}n_{\vb*{k}_1'}}{n_{\vb*{k}}n_{\vb*{k}_1}(1+n_{\vb*{k}'})(1+n_{\vb*{k}'_1})}\right]\nonumber\\
	&\times[(1+n_{\vb*{k}})(1+n_{\vb*{k}_1})n_{\vb*{k}'}n_{\vb*{k}_1'}-n_{\vb*{k}}n_{\vb*{k}_1}(1+n_{\vb*{k}'})(1+n_{\vb*{k}_1'})]\nonumber\\
	&\equiv\left.\pdv{H(\vb*{r},t)}{t}\right|_\mathrm{coll:N}.\label{S:Boltzmann3}
\end{align}
The integrand is of the form $(x-y)\log(x/y)$, and hence nonnegative. Thus, the H-function always decrease in the approach to equilibrium. This is the famaous Boltzmann's H-theorem.

The H-theorem shows that the entropy density, which for a Bose gas is given by
\begin{subequations}
	\begin{align}
		s(\vb*{r},t)&\equiv-k_\mathrm{B}H(\vb*{r},t)\\
		&=-k_\mathrm{B}\int[\dd{\vb*{k}}]\Bigl\{(1+n_{\vb*{k}})\log(1+n_{\vb*{k}})-n_{\vb*{k}}\log n_{\vb*{k}}\Bigr\},
	\end{align}
\end{subequations}
reaches a maximum in equilibrium. Here, $k_\mathrm{B}$ is Boltzmann's constant.

\subsection{III. Collision Invariants}
We will show that the Boltzmann equation Eq.~\eqref{S:Boltzmann} describes indeed an approach to the well-known Bose-Einstein equilibrium distribution. For this purpose, we first formalize the conservation laws. Here, we define the collision invariants $\mathcal{N}_\lambda$ as
\begin{equation}
	\int[\dd{\vb*{k}}]\mathcal{C}^\mathrm{mm}_\mathrm{N}[n_{\vb*{k}}]\mathcal{N}_\lambda=0.
\end{equation}
From this definition and Eq.~\eqref{S:Boltzmann2}, we immediately identify $\mathcal{N}_\lambda$ as,
\begin{equation}
	\mathcal{N}_0=1,\qquad\mathcal{N}_i=\hbar k_i,\qquad\mathcal{N}_4=\hbar\omega_{\vb*{k}}.\label{S:invariants}
\end{equation}
Multiplying the Boltzmann equation Eq.\eqref{S:Boltzmann} by the collision invariants $\mathcal{N}_\lambda$ gives the conservation laws for each invariant density respectively:
\begin{subequations}\label{S:Conservation}
	\begin{align}
		&\pdv{t}\int[\dd{\vb*{k}}]\mathcal{N}_\lambda n_{\vb*{k}}+\div\int[\dd{\vb*{k}}]\pdv{\omega_{\vb*{k}}}{\vb*{k}}\mathcal{N}_\lambda n_{\vb*{k}}=\pdv{t}\langle\mathcal{N}_\lambda\rangle+\div\vb*{\mathcal{J}}^{\mathcal{N}_\lambda}\\
		&\quad=\int[\dd{\vb*{k}}](\mathcal{C}^\mathrm{mm}_\mathrm{N}[n_{\vb*{k}}]+\mathcal{C}^\mathrm{mm}_\mathrm{U}[n_{\vb*{k}}])\mathcal{N}_\lambda=\int[\dd{\vb*{k}}]\mathcal{C}^\mathrm{mm}_\mathrm{U}[n_{\vb*{k}}]\mathcal{N}_\lambda=-\int[\dd{\vb*{k}}]\frac{n_{\vb*{k}}-N_{\vb*{k}}}{\tau_\mathrm{U}}\mathcal{N}_\lambda,  
	\end{align}
	where we have defined the invariant densities $\langle\mathcal{N}_\lambda\rangle$ and corresponding fluxes $\vb*{\mathcal{J}}^{\mathcal{N}_\lambda}$ as follows:
	\begin{equation}
		\langle\mathcal{N}_\lambda(\vb*{r},t)\rangle=\int[\dd{\vb*{k}}]\mathcal{N}_\lambda n_{\vb*{k}}(\vb*{r},t),\qquad\vb*{\mathcal{J}}^{\mathcal{N}_\lambda}(\vb*{r},t)=\int[\dd{\vb*{k}}]\pdv{\omega_{\vb*{k}}}{\vb*{k}}\mathcal{N}_\lambda n_{\vb*{k}}(\vb*{r},t).
	\end{equation}
\end{subequations}

\subsection{IV. Local Equilibrium Distribution}
The local equilibrium distribution $n_{\vb*{k}}^{(0)}$ is derived from the term in square brackets in Eq.~\eqref{S:Boltzmann3} has to vanish:
\begin{equation}
	\log\frac{n^{(0)}_{\vb*{k}}}{1+n^{(0)}_{\vb*{k}}}+\log\frac{n^{(0)}_{\vb*{k}_1}}{1+n^{(0)}_{\vb*{k}_1}}=\log\frac{n^{(0)}_{\vb*{k}'}}{1+n^{(0)}_{\vb*{k}'}}+\log\frac{n^{(0)}_{\vb*{k}'_1}}{1+n^{(0)}_{\vb*{k}'_1}}.
\end{equation}
This indicates that $\log[n^{(0)}/(1+n^{(0)})]$ is also a conserved quantity. Owing to basic conservation laws Eqs.~\eqref{S:Conservation} which we obtained, this quantity can be expressed as a linear combination of the collision invariants $\mathcal{N}_\lambda$:
\begin{equation}
	\log\frac{n^{(0)}_{\vb*{k}}}{1+n^{(0)}_{\vb*{k}}}=-\sum_\lambda\beta_\lambda\mathcal{N}_\lambda,\label{S:LED}
\end{equation}
with 
\begin{equation}
	\beta_0=\frac{-\tilde{\mu}}{k_\mathrm{B}T},\qquad\beta_i=\frac{-\vb*{v}}{k_\mathrm{B}T},\qquad\beta_4=\frac{1}{k_\mathrm{B}T},\label{S:Conjugate}
\end{equation}
where $\tilde{\mu}=\mu-m^\ast\vb*{v}^2/2$ is the magnon chemical potential in the frame which moves with fluid and $\vb*{v}$ is the drift velocity of the magnon system. All the expressions in Eq.~\eqref{S:Conjugate} can be slowly varying functions of $\vb*{r}$ and $t$. Such a situation is called a local equilibrium and Eq.~\eqref{S:LED} has the solution
\begin{align}
	n_{\vb*{k}}^{(0)}(\vb*{r},t)=n_\mathrm{B}\biggl(\sum_\lambda\beta_\lambda(\vb*{r},t)\mathcal{N}_\lambda\biggr)=\left[\exp\left\{\frac{\hbar\omega_{\vb*{k}}-\hbar\vb*{k}\vdot\vb*{v}(\vb*{r},t)-\tilde{\mu}(\vb*{r},t)}{k_\mathrm{B}T(\vb*{r},t)}\right\}-1\right]^{-1},
\end{align}
which is referred to as the local equilibrium distribution. Here, $n_{\mathrm{B}}(x)=1/(e^x-1)$ is the Bose-Einstein function and we use the quadratic dispersion for magnons $\hbar\omega_{\vb*{k}}=\hbar^2\vb*{k}^2/2m^\ast$ with the effective mass $m^\ast$. A similar derivation for the Boltzmann equation with U scattering process results in an equilibrium magnon distribution of the form
\begin{equation}
	\qquad  N_{\vb*{k}}(\vb*{r},t)=\left[\exp\left\{\frac{\hbar\omega_{\vb*{k}}-\mu(\vb*{r},t)}{k_\mathrm{B}T(\vb*{r},t)}\right\}-1\right]^{-1},
\end{equation}
because the drift velocity of magnons, whose total momentum is not conserved, is identical to zero.
\begin{comment}
	\begin{subequations}
		\begin{equation}
			\int[\dd{\vb*{k}}]n_{\vb*{k}}=\int[\dd{\vb*{k}}]n_{\vb*{k}}^{(0)}=\int[\dd{\vb*{k}}]N_{\vb*{k}}.
		\end{equation}
		Here, we have used the relations between distribution functions:
		\begin{equation}
			\int[\dd{\vb*{k}}]\hbar\vb*{k}n_{\vb*{k}}=\int[\dd{\vb*{k}}]\hbar\vb*{k}n^{(0)}_{\vb*{k}},\qquad\int[\dd{\vb*{k}}]\hbar\vb*{k}N_{\vb*{k}}=0,
		\end{equation}
	\end{subequations}
\end{comment}

\subsection{V. Hydrodynamic Equations}
We are now ready to derive the hydrodynamic equations from the conservation laws Eqs.~\eqref{S:Conservation}. In the following analysis, we assume that the local equilibrium distribution $n_{\vb*{k}}^{(0)}$ is transformed into the equilibrium distribution $N_{\vb*{k}}$ under the Galilean transformation from the frame in which the fluid moves with the velocity $\vb*{v}$ to the frame in which it is at rest: $\hbar\vb*{k}\to\hbar\vb*{k}+m^\ast\vb*{v}$.
\subsubsection{Number Conservation}
First, we start from the number conservation law by setting $\mathcal{N}_0=1$,
\begin{subequations}
	\begin{align}
		\langle\mathcal{N}_0\rangle&=\int[\dd{\vb*{k}}]n_{\vb*{k}}\equiv n(\vb*{r},t),\\
		\vb*{\mathcal{J}}^{\mathcal{N}_0}&=\int[\dd{\vb*{k}}]\pdv{\omega_{\vb*{k}}}{\vb*{k}}n_{\vb*{k}}=\int[\dd{\vb*{k}}]\left(\pdv{\omega_{\vb*{k}}}{\vb*{k}}-\vb*{v}\right)n_{\vb*{k}}+\vb*{v}\int[\dd{\vb*{k}}]n_{\vb*{k}}\nonumber\\
		&=\vb*{J}+n(\vb*{r},t)\vb*{v}(\vb*{r},t).\label{S:magnoncurrent}
	\end{align}
	Here, we have defined $\vb*{J}$ as
	\begin{equation}
		\vb*{J}\equiv\int[\dd{\vb*{k}}]\left(\pdv{\omega_{\vb*{k}}}{\vb*{k}}-\vb*{v}\right)n_{\vb*{k}},
	\end{equation}
	whose physical meaning becomes clear when we consider the entropy production.
\end{subequations}
After all, we obtain the number conservation law as follows:
\begin{equation}
	\pdv{n}{t}+\div(n\vb*{v}+\vb*{J})=0.\label{S:mass}
\end{equation}

\subsubsection{Momentum Conservation}
Next, we calculate the momentum conservation law by setting $\mathcal{N}_i=\hbar k_i$,
\begin{subequations}
	\begin{align}
		\langle\mathcal{N}_i\rangle&=\int[\dd{\vb*{k}}]\hbar k_in_{\vb*{k}}\equiv m^\ast nv_i\equiv\rho v_i,\\
		\mathcal{J}_j^{\mathcal{N}_i}&=\int[\dd{\vb*{k}}]\pdv{\omega_{\vb*{k}}}{k_j}\hbar k_in_{\vb*{k}}=\int[\dd{\vb*{k}}]\left(\pdv{\omega_{\vb*{k}}}{k_j}-v_j\right)\left(\hbar k_i-m^{\ast} v_i\right)n_{\vb*{k}}+\int[\dd{\vb*{k}}]\left(m^\ast\pdv{\omega_{\vb*{k}}}{k_j}v_i+v_j\hbar k_i\right)n_{\vb*{k}}-m^{\ast} v_jv_i\int[\dd{\vb*{k}}]n_{\vb*{k}}\nonumber\\
		&=\Pi_{ji}+m^\ast\mathcal{J}_j^{\mathcal{N}_0}v_i+v_j\langle\mathcal{N}_i\rangle -\rho v_jv_i=\Pi_{ji}+m^\ast\mathcal{J}_j^{\mathcal{N}_0}v_i,
	\end{align}
	where we have defined the stress tensor as
	\begin{equation}
		\Pi_{ji}\equiv\int[\dd{\vb*{k}}]\left(\pdv{\omega_{\vb*{k}}}{k_j}-v_j\right)\left(\hbar k_i-m^\ast v_i\right)n_{\vb*{k}}.
	\end{equation}
\end{subequations}
Therefore, the balance equation for momentum is obtained as
\begin{subequations}
	\begin{equation}
		\pdv{t}(\rho v_i)+\pdv{x_j}\left[(\rho v_j+J_j)v_i\right]+\pdv{x_j}\Pi_{ji}=-\frac{\rho v_i}{\tau_\mathrm{U}},\label{S:momentum1}
	\end{equation}
	or
	\begin{equation}
		\rho \pdv{v_i}{t}+(\rho v_j+J_j)\pdv{v_i}{x_j}+\pdv{x_j}\Pi_{ji}=-\frac{\rho_{}v_i}{\tau_\mathrm{U}},\label{S:momentum2}
	\end{equation}
\end{subequations}
with the help of Eq.~\eqref{S:mass}. Multiplying Eq.~\eqref{S:momentum2} by $v_i$ and summing over $i$, we obtain the following relation,
\begin{subequations}
	\begin{align}
		\rho v_i\pdv{v_i}{t}+(\rho v_j+J_j)v_i\pdv{v_i}{x_j}&=\pdv{t}\left(\frac{\rho }{2}v_iv_i\right)+\pdv{x_j}\left[\left(\rho v_j+J_j\right)\frac{v_iv_i}{2}\right]-\frac{v_iv_i}{2}\underbrace{\left[\pdv{\rho }{t}+\pdv{x_j}(\rho v_j+J_j)\right]}_{0}\nonumber\\
		&=\pdv{t}\left(\frac{\rho }{2}v_iv_i\right)+\pdv{x_j}\left[(\rho v_j+J_j)\frac{v_iv_i}{2}\right].
	\end{align}
	Therefore, we can express Eq.~\eqref{S:momentum2} in another form:
	\begin{equation}
		\pdv{t}\left(\frac{\rho }{2}v_iv_i\right)+\pdv{x_j}\left[(\rho v_j+J_j)\frac{v_iv_i}{2}\right]+v_i\pdv{x_j}\Pi_{ji}=-\frac{\rho  v_iv_i}{\tau_\mathrm{U}}.\label{S:momentum3}
	\end{equation}
\end{subequations}

\subsubsection{Energy conservation}
Finally, we calculate the energy conservation law by setting $\mathcal{N}_4=\hbar\omega_{\vb*{k}}$,
\begin{subequations}
	\begin{align}
		\langle\mathcal{N}_4\rangle&=\int[\dd{\vb*{k}}]\hbar\omega_{\vb*{k}}n_{\vb*{k}}=\int[\dd{\vb*{k}}]\frac{(\hbar\vb*{k}-m^\ast\vb*{v})^2}{2m}n_{\vb*{k}}+\int[\dd{\vb*{k}}]\hbar k_iv_in_{\vb*{k}}-\frac{m^\ast}{2}\vb*{v}^2\int[\dd{\vb*{k}}]n_{\vb*{k}}\nonumber\\
		&=u+\langle\mathcal{N}_i\rangle v_i-\frac{\rho }{2}\vb*{v}^2=u+\frac{\rho }{2}\vb*{v}^2,\\
		\mathcal{J}_j^{\mathcal{N}_4}&\equiv\int[\dd{\vb*{k}}]\pdv{\omega_{\vb*{k}}}{k_j}\hbar\omega_{\vb*{k}}n_{\vb*{k}}\nonumber\\
		&=\int[\dd{\vb*{k}}]\left(\pdv{\omega_{\vb*{k}}}{k_j}-v_j\right)\frac{(\hbar\vb*{k}-m^\ast\vb*{v})^2}{2m^\ast}n_{\vb*{k}}+\int[\dd{\vb*{k}}]\left(\pdv{\omega_{\vb*{k}}}{k_j}-v_j\right)(\hbar\vb*{k}-m^\ast\vb*{v})\vdot\vb*{v}n_{\vb*{k}}+\frac{m^\ast}{2}\vb*{v}^2\int[\dd{\vb*{k}}]\left(\pdv{\omega_{\vb*{k}}}{k_j}-v_j\right)n_{\vb*{k}}\nonumber\\
		&\quad+v_j\int[\dd{\vb*{k}}]\hbar\omega_{\vb*{k}}n_{\vb*{k}}\nonumber\\
		&=Q_j+\Pi_{ji}v_i+\frac{\vb*{v}^2}{2}(\mathcal{J}_j^{\mathcal{N}_0}-\rho v_j)+\langle\mathcal{N}_4\rangle v_j=Q_j+\Pi_{ji}v_i+\frac{\vb*{v}^2}{2}J_j+\left(u+\frac{\rho }{2}\vb*{v}^2\right)v_j,
	\end{align}
	where we have defined the internal energy density $u$ and the heat current $\vb*{Q}$ as
	\begin{align}
		u\equiv\int[\dd{\vb*{k}}]\frac{(\hbar\vb*{k}-m^\ast\vb*{v})^2}{2m^\ast}n_{\vb*{k}},\qquad
		\vb*{Q}\equiv\int[\dd{\vb*{k}}]\left(\pdv{\omega_{\vb*{k}}}{\vb*{k}}-\vb*{v}\right)\frac{(\hbar\vb*{k}-m^\ast\vb*{v})^2}{2m^\ast}n_{\vb*{k}}.\label{S:heatcurrent}
	\end{align}
\end{subequations}
Therefore, the energy conservation law becomes
\begin{equation}
	\pdv{t}\left(u+\frac{\rho }{2}\vb*{v}^2\right)+\pdv{x_j}\left[uv_j+Q_j+(\rho v_j+J_j)\frac{\vb*{v}^2}{2}\right]+\pdv{x_j}(\Pi_{ji}v_i)=0,\label{S:energy1}
\end{equation}
or
\begin{equation}
	\pdv{u}{t}+\div(u\vb*{v}+\vb*{Q})+\Pi_{ji}\pdv{x_j}v_i=\frac{\rho v_iv_i}{\tau_\mathrm{U}},\label{S:energy2}
\end{equation}
by employing Eq.~\eqref{S:momentum3}.

\subsection{VI. Deviation from the Local Equilibrium Distribution}
By considering the deviation from the local equilibrium distribution as $n_{\vb*{k}}'=n_{\vb*{k}}-n_{\vb*{k}}^{(0)}$, the fluxes for each conserved quantity are devided into two contribution:
\begin{subequations}
	\begin{equation}
		\vb*{J}=\vb*{J}^{(0)}+\vb*{J}',\qquad\Pi_{ji}=\Pi^{(0)}_{ji}+\Pi_{ji}',\qquad\vb*{Q}=\vb*{Q}^{(0)}+\vb*{Q}',
	\end{equation}
	with
	\begin{align}
		\vb*{J}^{(0)}&=\int[\dd{\vb*{k}}]\left(\pdv{\omega_{\vb*{k}}}{\vb*{k}}-\vb*{v}\right)n_{\vb*{k}}^{(0)}=\int[\dd{\vb*{k}}]\pdv{\omega_{\vb*{k}}}{\vb*{k}}N_{\vb*{k}}=0,\\
		\Pi_{ji}^{(0)}&=\int[\dd{\vb*{k}}]\left(\pdv{\omega_{\vb*{k}}}{k_j}-v_j\right)\left(\hbar k_i-m^\ast v_i\right)n_{\vb*{k}}^{(0)}=\int[\dd{\vb*{k}}]\pdv{\omega_{\vb*{k}}}{k_j}\hbar k_iN_{\vb*{k}}=P\delta_{ij},\\
		\vb*{Q}^{(0)}&=\int[\dd{\vb*{k}}]\left(\pdv{\omega_{\vb*{k}}}{\vb*{k}}-\vb*{v}\right)\frac{(\hbar\vb*{k}-m^\ast\vb*{v})^2}{2m^\ast}n^{(0)}_{\vb*{k}}=\int[\dd{\vb*{k}}]\pdv{\omega_{\vb*{k}}}{\vb*{k}}\frac{(\hbar\vb*{k})^2}{2m^\ast}N_{\vb*{k}}=0,\label{S:heat-zeroth}
	\end{align}
	where we have introduced the pressure $P$ as
	\begin{equation}
		P\equiv-k_\mathrm{B}T\int[\dd{\vb*{k}}]\log[1-e^{-\beta(\hbar\omega_{\vb*{k}}-\mu)}].
	\end{equation}
\end{subequations}
By using these quantities, the hydrodynamic equations Eqs.~\eqref{S:momentum3} and ~\eqref{S:energy2} are rewritten as
\begin{align}
	\rho \pdv{v_i}{t}+(\rho v_j+J_j')\pdv{v_i}{x_j}+\pdv{x_i}P+\pdv{x_j}\Pi_{ji}'=-\frac{\rho v_i}{\tau_\mathrm{U}},\\
	\pdv{u}{t}+\div(u\vb*{v})+P\div\vb*{v}+\div\vb*{Q}'+\Pi'_{ji}\pdv{v_i}{x_j}=\frac{\rho \vb*{v}^2}{\tau_\mathrm{U}}.\label{S:momentum4}
\end{align}
Comparing with the phenomelogical hydrodynamics, $\vb*{J}'$, $\Pi_{ji}'$, and $\vb{Q}'$ correspond to the diffusion current, the viscous stress, and the heat current respectively. Namely, dissipative currents emerge only when considering the deviation from the local equilibrium.

\section{Entropy production}
By employing nonequilibrium thermodynamics, we first assume that the Gibbs-Duhem equation
\begin{equation}
	Ts=u+P-\mu n ,\label{GD:1}
\end{equation}
and the thermodynamic relation
\begin{equation}
	s\dd{T}=\dd{P}-n \dd{\mu}.\label{S:thermo_relation}
\end{equation}
Then, we rewrite Eq.~\eqref{GD:1} by using our invariant densities as
\begin{equation}
	Ts=\langle\mathcal{N}_4\rangle-\langle\mathcal{N}_i\rangle v_i+P-\tilde{\mu}\langle\mathcal{N}_0\rangle.\label{GD:2}
\end{equation}
From Eqs.~\eqref{S:thermo_relation} and ~\eqref{GD:2}, we obtain
\begin{align}
	T\dd{s}&=\dd{u}-\mu\dd{n}\nonumber\\
	&=\dd{\langle\mathcal{N}_4\rangle}-v_i\dd{\langle\mathcal{N}_i\rangle}
	-\tilde{\mu}\dd{\langle\mathcal{N}_0\rangle}.
\end{align}
From this equation, we obtain the following relations,
\begin{subequations}
	\begin{align}
		\pdv{s}{t}&=\frac{1}{T}\pdv{\langle\mathcal{N}_4\rangle}{t}-\frac{v_i}{T}\pdv{\langle\mathcal{N}_i\rangle}{t}-\frac{\tilde{\mu}}{T}\pdv{\mathcal{N}_0}{t},\\
		\grad s&=\frac{1}{T}\grad\langle\mathcal{N}_4\rangle-\frac{v_i}{T}\grad\langle\mathcal{N}_i\rangle-\frac{\tilde{\mu}}{T}\grad\mathcal{N}_0,
	\end{align}
\end{subequations}
By using the conservation laws we derived, we construct the entropy production equation as follows:
\begin{align}
	&\pdv{s}{t}+\div(s\vb*{v})=\pdv{s}{t}+s\div\vb*{v}+\vb*{v}\vdot\grad s\nonumber\\
	&=-\div\frac{(\vb*{\mathcal{J}}^{\mathcal{N}_4}-\langle\mathcal{N}_4\rangle\vb*{v})-v_i(\vb*{\mathcal{J}}^{\mathcal{N}_i}-\langle\mathcal{N}_i\rangle\vb*{v})-\tilde{\mu}(\vb*{\mathcal{J}}^{\mathcal{N}_0}-\langle\mathcal{N}_0\rangle\vb*{v})}{T}\nonumber\\
	&\quad+(\vb*{\mathcal{J}}^{\mathcal{N}_4}-\langle\mathcal{N}_4\rangle\vb*{v})\vdot\grad\left(\frac{1}{T}\right)+\frac{P}{T}(\div\vb*{v})+(\vb*{\mathcal{J}}^{\mathcal{N}_i}-\langle\mathcal{N}_i\rangle\vb*{v})\vdot\grad\left(-\frac{v_i}{T}\right)+(\vb*{\mathcal{J}}^{\mathcal{N}_0}-\langle\mathcal{N}_0\rangle\vb*{v})\vdot\grad\left(-\frac{\tilde{\mu}}{T}\right).\label{EP:1}
\end{align}
Substituting the form of hydrodynamic variables into each invariant density and corresponding flux, the entropy production rate becomes
\begin{subequations}
	\begin{align}
		&\pdv{s}{t}+\div\left(s\vb*{v}+\frac{\vb*{Q}'-{\mu}\vb*{J}'}{T}\right)\nonumber\\
		&=\vb*{Q}'\vdot\grad\left(\frac{1}{T}\right)+\frac{\Pi_{ji}'}{T}\partial_j(-v_i)+\vb*{J}'\vdot\grad\left(-\frac{\mu}{T}\right)+\frac{1}{T}\frac{\rho \vb*{v}^2}{\tau_\mathrm{U}},\label{EP:2}\\
		&=\frac{\vb*{Q}'-{\mu}\vb*{J}'}{T}\vdot\left(-\frac{\grad T}{T}\right)+\frac{\Pi_{ji}'}{T}\grad_j(-v_i)+\frac{\vb*{J}'}{T}\vdot\grad(-{\mu})+\frac{1}{T}\frac{\rho \vb*{v}^2}{\tau_\mathrm{U}}.\label{EP:3}
	\end{align}
\end{subequations}
As can be seen, the entropy is irreversively produced by the dissipation currents stem from the deviation from the local equilibrium. The entropy production rate Eq.~\eqref{EP:3} is given by the product of dissipation currents and corresponding thermodynamic forces:
\begin{equation}
	\left(\pdv{s}{t}\right)_{\mathrm{irr}}=\sum_{i}Y_iX_i.
\end{equation}
If we write the response of the general currents $Y_i$ induced by such thermodynamic forces $X_i$ as
\begin{equation}
	Y_i=\sum_j\mathcal{L}_{ij}X_i,
\end{equation}
the response coefficients $\mathcal{L}_{ij}$ becomes symmetric $\mathcal{L}_{ij}=\mathcal{L}_{ji}$, which is well-known as the Onsager's reciprocal relations.

\section{Heat current}
In this section, we derive the heat current by considering deviation from the local equilibrium distribution $n_{\vb*{k}}^{(0)}$.
In the following analysis, we also assume the relaxation approximation for N process:
\begin{equation}
	C_\mathrm{N}^\mathrm{mm}[n_{\vb*{k}}]=-\frac{n_{\vb*{k}}(\vb*{r},t)-n^{(0)}_{\vb*{k}}(\vb*{r},t)}{\tau_\mathrm{N}}.
\end{equation}
For the purpose of obtaining the spin and thermal conductivities, we expand the magnon distribution function around local equilibrium: $n_{\vb*{k}}=n_{\vb*{k}}^{(0)}+\delta n_{\vb*{k}}$. Assuming that the deviation from local equilibrium is small to justify the use of the linearized Boltzmann equation, we can calculate $\delta n_{\vb*{k}}$ by neglecting its time and spacial derivatives,
\begin{subequations}
	\begin{align}
		&\left(\pdv{t}+\pdv{\omega_{\vb*{k}}}{\vb*{k}}\vdot\grad\right)n_{\vb*{k}}\simeq\left(\pdv{t}+\pdv{\omega_{\vb*{k}}}{\vb*{k}}\vdot\grad\right)n_{\vb*{k}}^{(0)}\nonumber\\
		&\quad=-\frac{\delta n_{\vb*{k}}}{\tau_\mathrm{N}}-\frac{n_{\vb*{k}}^{(0)}+\delta n_{\vb*{k}}-N_{\vb*{k}}}{\tau_\mathrm{U}}=-\left(\frac{1}{\tau_\mathrm{N}}+\frac{1}{\tau_\mathrm{U}}\right)\delta n_{\vb*{k}}-\frac{n_{\vb*{k}}^{(0)}-N_{\vb*{k}}}{\tau_\mathrm{U}}.
	\end{align}
	Therefore, we obtain the deviation $\delta n_{\vb*{k}}$ in the first approximation as,
	\begin{equation}
		\delta n_{\vb*{k}}\simeq-\left(\frac{1}{\tau_\mathrm{N}}+\frac{1}{\tau_\mathrm{U}}\right)^{-1}\left[\left(\pdv{t}+\pdv{\omega_{\vb*{k}}}{\vb*{k}}\vdot\grad\right)n_{\vb*{k}}^{(0)}+\frac{n_{\vb*{k}}^{(0)}-N_{\vb*{k}}}{\tau_\mathrm{U}}\right].\label{S:deviation1}
	\end{equation}
\end{subequations}
Furthermore, we assume that $\delta n_{\vb*{k}}$ satisfies the following condition,
\begin{subequations}
	\begin{equation}
		\int[\dd{\vb*{k}}]\mathcal{N}_\lambda\delta n_{\vb*{k}}=0,
	\end{equation}
	which results in the fact that the invariant densities is determined only by the local distribution function:
	\begin{equation}
		\langle\mathcal{N}_\lambda\rangle=\int[\dd{\vb*{k}}]\mathcal{N}_\lambda(n_{\vb*{k}}^{(0)}+\delta n_{\vb*{k}})=\int[\dd{\vb*{k}}]\mathcal{N}_\lambda n_{\vb*{k}}^{(0)}=\langle\mathcal{N}_\lambda\rangle^{(0)}.
	\end{equation}
\end{subequations}
This condition is necessary for not violating the conservation laws under the relaxation time approximation for N process. In the following analysis, we restrict ourselves to steady states and therefore neglect the time derivative $\partial/\partial t$ in Eq.~\eqref{S:deviation1}.

By substituting Eq.~\eqref{S:deviation1} into Eq.~\eqref{S:heatcurrent}, we can calculate the heat current as,
\begin{subequations}
	\begin{align}
		\vb*{Q}&=\int[\dd{\vb*{k}}]\left(\pdv{\omega_{\vb*{k}}}{\vb*{k}}-\vb*{v}\right)\frac{(\hbar\vb*{k}-m^\ast\vb*{v})^2}{2m^\ast}\delta n_{\vb*{k}}\nonumber\\
		&\simeq-\left(\frac{1}{\tau_\mathrm{N}}+\frac{1}{\tau_\mathrm{U}}\right)^{-1}\int[\dd{\vb*{k}}]\left(\pdv{\omega_{\vb*{k}}}{\vb*{k}}-\vb*{v}\right)\frac{(\hbar\vb*{k}-m^\ast\vb*{v})^2}{2m^\ast}\left[\pdv{\omega_{\vb*{k}}}{\vb*{k}}\vdot\grad n_{\vb*{k}}^{(0)}+\frac{n_{\vb*{k}}^{(0)}-N_{\vb*{k}}}{\tau_\mathrm{U}}\right]\nonumber\\
		&\equiv\vb*{Q}^{\grad}+\vb*{Q}^{\tau}.
	\end{align}
	Here we have defined $\vb*{Q}^{\grad}$ and $\vb*{Q}^{\tau}$ as
	\begin{align}
		\vb*{Q}^{\grad}&\equiv-\left(\frac{1}{\tau_\mathrm{N}}+\frac{1}{\tau_\mathrm{U}}\right)^{-1}\int[\dd{\vb*{k}}]\left(\pdv{\omega_{\vb*{k}}}{\vb*{k}}-\vb*{v}\right)\frac{(\hbar\vb*{k}-m^\ast\vb*{v})^2}{2m^\ast}\pdv{\omega_{\vb*{k}}}{\vb*{k}}\vdot\grad n_{\vb*{k}}^{(0)},\\
		\vb*{Q}^{\tau}&\equiv-\left(\frac{1}{\tau_\mathrm{N}}+\frac{1}{\tau_\mathrm{U}}\right)^{-1}\int[\dd{\vb*{k}}]\left(\pdv{\omega_{\vb*{k}}}{\vb*{k}}-\vb*{v}\right)\frac{(\hbar\vb*{k}-m^\ast\vb*{v})^2}{2m^\ast}\frac{n_{\vb*{k}}^{(0)}-N_{\vb*{k}}}{\tau_\mathrm{U}}.
	\end{align}
\end{subequations}
The contribution from U process $\vb*{Q}^\tau$ can be easily calculated by using the relation Eq.~\eqref{S:heat-zeroth} as,
\begin{align}
	\vb*{Q}^{\tau}&=\frac{1}{\tau_\mathrm{U}}\left(\frac{1}{\tau_\mathrm{N}}+\frac{1}{\tau_\mathrm{U}}\right)^{-1}\int[\dd{\vb*{k}}]\left(\pdv{\omega_{\vb*{k}}}{\vb*{k}}-\vb*{v}\right)\frac{(\hbar\vb*{k}-m^\ast\vb*{v})^2}{2m^\ast}N_{\vb*{k}}\nonumber\\
	&=\frac{-1}{\tau_\mathrm{U}}\left(\frac{1}{\tau_\mathrm{N}}+\frac{1}{\tau_\mathrm{U}}\right)^{-1}\int[\dd{\vb*{k}}]\left\{\frac{(\hbar\vb*{k})^2}{2m^\ast}\vb*{v}+{\pdv{\omega_{\vb*{k}}}{\vb*{k}}}(\hbar\vb*{k}\vdot\vb*{v})+\left(\frac{m^\ast}{2}\vb*{v}^2\right)\vb*{v}\right\}N_{\vb*{k}}\nonumber\\
	&=\frac{-1}{\tau_\mathrm{U}}\left(\frac{1}{\tau_\mathrm{N}}+\frac{1}{\tau_\mathrm{U}}\right)^{-1}\left(u^{(0)}+P+\frac{\rho ^{(0)}}{2}\vb*{v}^2\right)\vb*{v}.\label{S:heattau}
\end{align}
On the other hand, the contribution from spatial gradients $\vb*{Q}^{\grad}$ needs to identify the spatial dependence of $n_{\vb*{k}}^{(0)}$. $n_{\vb*{k}}^{(0)}$ depends on space through $\tilde{\mu}$, $\vb*{v}$, and $T$, thus $\grad n_{\vb*{k}}^{(0)}$ can be calculated as,
\begin{align}
	\grad n_{\vb*{k}}^{(0)}&=\sum_\lambda\pdv{n_{\vb*{k}}^{(0)}}{\beta_\lambda}\grad\beta_\lambda\nonumber\\
	&=\biggl[\sum_\lambda\mathcal{N}_\lambda\grad\beta_\lambda\biggr]\left.\pdv{n_\mathrm{B}(x)}{x}\right|_{x=\sum_\lambda\beta_\lambda\mathcal{N}_\lambda}\nonumber\\
	&=T\left[\hbar\omega_{\vb*{k}}\grad\left(\frac{1}{T}\right)+\hbar k_j\grad\left(-\frac{v_j}{T}\right)+\grad\left(-\frac{\tilde{\mu}}{T}\right)\right]\pdv{n_{\vb*{k}}^{(0)}}{(\hbar\omega_{\vb*{k}})}.
\end{align}
Therefore, $\vb*{Q}^{\grad}$ is further decomposed into three contributions: $\vb*{Q}^{\grad}=\vb*{Q}^{\grad T}+\vb*{Q}^{\grad\vb*{v}}+\vb*{Q}^{\grad\mu}$,
\begin{subequations}\label{S:heat:grad}
	\begin{align}
		\vb*{Q}^{\grad T}&=-\left(\frac{1}{\tau_\mathrm{N}}+\frac{1}{\tau_\mathrm{U}}\right)^{-1}T\int[\dd{\vb*{k}}]\left(\pdv{\omega_{\vb*{k}}}{\vb*{k}}-\vb*{v}\right)\frac{(\hbar\vb*{k}-m^\ast\vb*{v})^2}{2m^\ast}\hbar\omega_{\vb*{k}}\left\{\pdv{\omega_{\vb*{k}}}{\vb*{k}}\vdot\grad\left(\frac{1}{T}\right)\right\}\pdv{n_{\vb*{k}}^{(0)}}{(\hbar\omega_{\vb*{k}})}\nonumber\\
		&=-\left(\frac{1}{\tau_\mathrm{N}}+\frac{1}{\tau_\mathrm{U}}\right)^{-1}T\int[\dd{\vb*{k}}]\left(\pdv{\omega_{\vb*{k}}}{\vb*{k}}-\vb*{v}\right)\frac{(\hbar\vb*{k}-m^\ast\vb*{v})^2}{2m^\ast}\frac{(\hbar\vb*{k}-m^\ast\vb*{v})^2}{2m^\ast}\left\{\left(\pdv{\omega_{\vb*{k}}}{\vb*{k}}-\vb*{v}\right)\vdot\grad\left(\frac{1}{T}\right)\right\}\pdv{n_{\vb*{k}}^{(0)}}{(\hbar\omega_{\vb*{k}})}\nonumber\\
		&\qquad-\left(\frac{1}{\tau_\mathrm{N}}+\frac{1}{\tau_\mathrm{U}}\right)^{-1}T\int[\dd{\vb*{k}}]\left(\pdv{\omega_{\vb*{k}}}{\vb*{k}}-\vb*{v}\right)\frac{(\hbar\vb*{k}-m^\ast\vb*{v})^2}{2m^\ast}\frac{(\hbar\vb*{k}-m^\ast\vb*{v})^2}{2m^\ast}\left\{\vb*{v}\vdot\grad\left(\frac{1}{T}\right)\right\}\pdv{n_{\vb*{k}}^{(0)}}{(\hbar\omega_{\vb*{k}})}\nonumber\\
		&\qquad-\left(\frac{1}{\tau_\mathrm{N}}+\frac{1}{\tau_\mathrm{U}}\right)^{-1}T\int[\dd{\vb*{k}}]\left(\pdv{\omega_{\vb*{k}}}{\vb*{k}}-\vb*{v}\right)\frac{(\hbar\vb*{k}-m^\ast\vb*{v})^2}{2m^\ast}\left(\hbar\vb*{k}-m^\ast\vb*{v}\right)\vdot\vb*{v}\left\{\left(\pdv{\omega_{\vb*{k}}}{\vb*{k}}-\vb*{v}\right)\vdot\grad\left(\frac{1}{T}\right)\right\}\pdv{n_{\vb*{k}}^{(0)}}{(\hbar\omega_{\vb*{k}})}\nonumber\\
		&\qquad-\left(\frac{1}{\tau_\mathrm{N}}+\frac{1}{\tau_\mathrm{U}}\right)^{-1}T\int[\dd{\vb*{k}}]\left(\pdv{\omega_{\vb*{k}}}{\vb*{k}}-\vb*{v}\right)\frac{(\hbar\vb*{k}-m^\ast\vb*{v})^2}{2m^\ast}\left(\hbar\vb*{k}-m^\ast\vb*{v}\right)\vdot\vb*{v}\left\{\vb*{v}\vdot\grad\left(\frac{1}{T}\right)\right\}\pdv{n_{\vb*{k}}^{(0)}}{(\hbar\omega_{\vb*{k}})}\nonumber\\
		&\qquad-\left(\frac{1}{\tau_\mathrm{N}}+\frac{1}{\tau_\mathrm{U}}\right)^{-1}T\int[\dd{\vb*{k}}]\left(\pdv{\omega_{\vb*{k}}}{\vb*{k}}-\vb*{v}\right)\frac{(\hbar\vb*{k}-m^\ast\vb*{v})^2}{2m^\ast}\frac{m^\ast}{2}\vb*{v}^2\left\{\left(\pdv{\omega_{\vb*{k}}}{\vb*{k}}-\vb*{v}\right)\vdot\grad\left(\frac{1}{T}\right)\right\}\pdv{n_{\vb*{k}}^{(0)}}{(\hbar\omega_{\vb*{k}})}\nonumber\\
		&\qquad-\left(\frac{1}{\tau_\mathrm{N}}+\frac{1}{\tau_\mathrm{U}}\right)^{-1}T\int[\dd{\vb*{k}}]\left(\pdv{\omega_{\vb*{k}}}{\vb*{k}}-\vb*{v}\right)\frac{(\hbar\vb*{k}-m^\ast\vb*{v})^2}{2m^\ast}\frac{m^\ast}{2}\vb*{v}^2\left\{\vb*{v}\vdot\grad\left(\frac{1}{T}\right)\right\}\pdv{n_{\vb*{k}}^{(0)}}{(\hbar\omega_{\vb*{k}})}.\label{S:gradT}\\
		\nonumber\\
		\vb*{Q}^{\grad\vb*{v}}&=-\left(\frac{1}{\tau_\mathrm{N}}+\frac{1}{\tau_\mathrm{U}}\right)^{-1}T\int[\dd{\vb*{k}}]\left(\pdv{\omega_{\vb*{k}}}{\vb*{k}}-\vb*{v}\right)\frac{(\hbar\vb*{k}-m^\ast\vb*{v})^2}{2m^\ast}\left\{\pdv{\omega_{\vb*{k}}}{\vb*{k}}\vdot\grad\left(- \frac{v_j}{T}\right)\right\}\hbar k_j\pdv{n_{\vb*{k}}^{(0)}}{(\hbar\omega_{\vb*{k}})}\nonumber\\
		&=-\left(\frac{1}{\tau_\mathrm{N}}+\frac{1}{\tau_\mathrm{U}}\right)^{-1}T\int[\dd{\vb*{k}}]\left(\pdv{\omega_{\vb*{k}}}{\vb*{k}}-\vb*{v}\right)\frac{(\hbar\vb*{k}-m^\ast\vb*{v})^2}{2m^\ast}\left\{\left(\pdv{\omega_{\vb*{k}}}{\vb*{k}}-\vb*{v}\right)\vdot\grad\left(-\frac{v_j}{T}\right)\right\}(\hbar k_j-m^\ast v_j)\pdv{n_{\vb*{k}}^{(0)}}{(\hbar\omega_{\vb*{k}})}\nonumber\\
		&\qquad-\left(\frac{1}{\tau_\mathrm{N}}+\frac{1}{\tau_\mathrm{U}}\right)^{-1}T\int[\dd{\vb*{k}}]\left(\pdv{\omega_{\vb*{k}}}{\vb*{k}}-\vb*{v}\right)\frac{(\hbar\vb*{k}-m^\ast\vb*{v})^2}{2m^\ast}\left\{\vb*{v}\vdot\grad\left(-\frac{v_j}{T}\right)\right\}(\hbar k_j-m^\ast v_j)\pdv{n_{\vb*{k}}^{(0)}}{(\hbar\omega_{\vb*{k}})}\nonumber\\
		&\qquad-\left(\frac{1}{\tau_\mathrm{N}}+\frac{1}{\tau_\mathrm{U}}\right)^{-1}T\int[\dd{\vb*{k}}]\left(\pdv{\omega_{\vb*{k}}}{\vb*{k}}-\vb*{v}\right)\frac{(\hbar\vb*{k}-m^\ast\vb*{v})^2}{2m^\ast}\left\{\left(\pdv{\omega_{\vb*{k}}}{\vb*{k}}-\vb*{v}\right)\vdot\grad\left( -\frac{v_j}{T}\right)\right\}m^\ast v_j\pdv{n_{\vb*{k}}^{(0)}}{(\hbar\omega_{\vb*{k}})}\nonumber\\
		&\qquad-\left(\frac{1}{\tau_\mathrm{N}}+\frac{1}{\tau_\mathrm{U}}\right)^{-1}T\int[\dd{\vb*{k}}]\left(\pdv{\omega_{\vb*{k}}}{\vb*{k}}-\vb*{v}\right)\frac{(\hbar\vb*{k}-m^\ast\vb*{v})^2}{2m^\ast}\left\{\vb*{v}\vdot\grad\left(-\frac{v_j}{T}\right)\right\}m^\ast v_j\pdv{n_{\vb*{k}}^{(0)}}{(\hbar\omega_{\vb*{k}})},\label{S:gradv1}\\
		\nonumber\\
		\vb*{Q}^{\grad\mu}&=-\left(\frac{1}{\tau_\mathrm{N}}+\frac{1}{\tau_\mathrm{U}}\right)^{-1}T\int[\dd{\vb*{k}}]\left(\pdv{\omega_{\vb*{k}}}{\vb*{k}}-\vb*{v}\right)\frac{(\hbar\vb*{k}-m^\ast\vb*{v})^2}{2m^\ast}\left\{\pdv{\omega_{\vb*{k}}}{\vb*{k}}\vdot\grad\left(-\frac{\tilde{\mu}}{T}\right)\right\}\pdv{n_{\vb*{k}}^{(0)}}{(\hbar\omega_{\vb*{k}})}\nonumber\\
		&=-\left(\frac{1}{\tau_\mathrm{N}}+\frac{1}{\tau_\mathrm{U}}\right)^{-1}T\int[\dd{\vb*{k}}]\left(\pdv{\omega_{\vb*{k}}}{\vb*{k}}-\vb*{v}\right)\frac{(\hbar\vb*{k}-m^\ast\vb*{v})^2}{2m^\ast}\left\{\left(\pdv{\omega_{\vb*{k}}}{\vb*{k}}-\vb*{v}\right)\vdot\grad\left(-\frac{\tilde{\mu}}{T}\right)\right\}\pdv{n_{\vb*{k}}^{(0)}}{(\hbar\omega_{\vb*{k}})}\nonumber\\
		&\qquad-\left(\frac{1}{\tau_\mathrm{N}}+\frac{1}{\tau_\mathrm{U}}\right)^{-1}T\int[\dd{\vb*{k}}]\left(\pdv{\omega_{\vb*{k}}}{\vb*{k}}-\vb*{v}\right)\frac{(\hbar\vb*{k}-m^\ast\vb*{v})^2}{2m^\ast}\left\{\vb*{v}\vdot\grad\left(-\frac{\tilde{\mu}}{T}\right)\right\}\pdv{n_{\vb*{k}}^{(0)}}{(\hbar\omega_{\vb*{k}})}.\label{S:gradmu1}
	\end{align}
\end{subequations}
By employing the Galilei transformation $\hbar\vb*{k}\to\hbar\vb*{k}+m^\ast\vb*{v}$: $n_{\vb*{k}+m^\ast\vb*{v}/\hbar}^{(0)}=N_{\vb*{k}}$, odd functions with respect to $\vb*{k}$ in Eqs.~\eqref{S:heat:grad} vanish and yield the following results,
\begin{subequations}
	\begin{align}
		\vb*{Q}^{\grad T}&=-\left(\frac{1}{\tau_\mathrm{N}}+\frac{1}{\tau_\mathrm{U}}\right)^{-1}T\int[\dd{\vb*{k}}]\pdv{\omega_{\vb*{k}}}{\vb*{k}}\frac{(\hbar\vb*{k})^2}{2m^\ast}\frac{(\hbar\vb*{k})^2}{2m^\ast}\left\{\pdv{\omega_{\vb*{k}}}{\vb*{k}}\vdot\grad\left(\frac{1}{T}\right)\right\}\pdv{N_{\vb*{k}}}{(\hbar\omega_{\vb*{k}})}\nonumber\\
		&\qquad-\left(\frac{1}{\tau_\mathrm{N}}+\frac{1}{\tau_\mathrm{U}}\right)^{-1}T\int[\dd{\vb*{k}}]\pdv{\omega_{\vb*{k}}}{\vb*{k}}\frac{(\hbar\vb*{k})^2}{2m^\ast}(\vb*{v}\vdot\hbar\vb*{k})\left\{(\vb*{v}\vdot\grad)\left(\frac{1}{T}\right)\right\}\pdv{N_{\vb*{k}}}{(\hbar\omega_{\vb*{k}})}\nonumber\\
		&\qquad-\left(\frac{1}{\tau_\mathrm{N}}+\frac{1}{\tau_\mathrm{U}}\right)^{-1}T\int[\dd{\vb*{k}}]\pdv{\omega_{\vb*{k}}}{\vb*{k}}\frac{(\hbar\vb*{k})^2}{2m^\ast}\left(\frac{m^\ast}{2}\vb*{v}^2\right)\left\{\pdv{\omega_{\vb*{k}}}{\vb*{k}}\vdot\grad\left(\frac{1}{T}\right)\right\}\pdv{N_{\vb*{k}}}{(\hbar\omega_{\vb*{k}})},\\
		\nonumber\\
		\vb*{Q}^{\grad\vb*{v}}&=-\left(\frac{1}{\tau_\mathrm{N}}+\frac{1}{\tau_\mathrm{U}}\right)^{-1}T\int[\dd{\vb*{k}}]\pdv{\omega_{\vb*{k}}}{\vb*{k}}\frac{(\hbar\vb*{k})^2}{2m^\ast}\left\{\vb*{v}\vdot\grad\left(-\frac{v_j}{T}\right)\right\}\hbar k_j\pdv{N_{\vb*{k}}}{(\hbar\omega_{\vb*{k}})}\nonumber\\
		&\qquad-\left(\frac{1}{\tau_\mathrm{N}}+\frac{1}{\tau_\mathrm{U}}\right)^{-1}T\int[\dd{\vb*{k}}]\pdv{\omega_{\vb*{k}}}{\vb*{k}}\frac{(\hbar\vb*{k})^2}{2m^\ast}\left\{\pdv{\omega_{\vb*{k}}}{\vb*{k}}\vdot\grad\left(-\frac{v_j}{T}\right)\right\}m^\ast v_j\pdv{N_{\vb*{k}}}{(\hbar\omega_{\vb*{k}})},\\
		\nonumber\\
		\vb*{Q}^{\grad\mu}&=-\left(\frac{1}{\tau_\mathrm{N}}+\frac{1}{\tau_\mathrm{U}}\right)^{-1}T\int[\dd{\vb*{k}}]\pdv{\omega_{\vb*{k}}}{\vb*{k}}\frac{(\hbar\vb*{k})^2}{2m^\ast}\left\{\pdv{\omega_{\vb*{k}}}{\vb*{k}}\vdot\grad\left(-\frac{\tilde{\mu}}{T}\right)\right\}\pdv{N_{\vb*{k}}}{(\hbar\omega_{\vb*{k}})}.
	\end{align}
\end{subequations}
By employing the following formula,
\begin{subequations}\label{S:formula}
	\begin{align}
		&\int[\dd{\vb*{k}}]\pdv{\omega_{\vb*{k}}}{k_i}\frac{(\hbar\vb*{k})^2}{2m^\ast}\pdv{\omega_{\vb*{k}}}{k_j}\pdv{N_{\vb*{k}}}{(\hbar\omega_{\vb*{k}})}=-\frac{u^{(0)}+P}{m^\ast}\delta_{ij},\\
		&\int[\dd{\vb*{k}}]\pdv{\omega_{\vb*{k}}}{k_i}\frac{(\hbar\vb*{k})^2}{2m^\ast}\pdv{\omega_{\vb*{k}}}{k_j}\hbar\omega_{\vb*{k}}\pdv{N_{\vb*{k}}}{(\hbar\omega_{\vb*{k}})}=-\left(\frac{4}{d}+1\right)\langle(\mathcal{N}_4)^2\rangle_\mathrm{eq}\frac{\delta_{ij}}{m^\ast},\qquad d=2,3
	\end{align}
\end{subequations}
we finally obtain the hydrodynamic expression for the heat current:
\begin{subequations}\label{S:grad}
	\begin{align}
		\vb*{Q}^{\grad T}&=\left(\frac{1}{\tau_\mathrm{N}}+\frac{1}{\tau_\mathrm{U}}\right)^{-1}T\left[\left(\frac{4}{d}+1\right)\frac{\langle(\mathcal{N}_4)^2\rangle_\mathrm{eq}}{m^\ast}+(u^{(0)}+P)\frac{\vb*{v}^2}{2}\right]\grad\left(\frac{1}{T}\right)\nonumber\\
		&\qquad+\left(\frac{1}{\tau_\mathrm{N}}+\frac{1}{\tau_\mathrm{U}}\right)^{-1}T(u^{(0)}+P)\vb*{v}(\vb*{v}\vdot\grad)\left(\frac{1}{T}\right),\label{S:gradT2}\\
		\vb*{Q}^{\grad\vb*{v}}&=\left(\frac{1}{\tau_\mathrm{N}}+\frac{1}{\tau_\mathrm{U}}\right)^{-1}T(u^{(0)}+P)\left[(\vb*{v}\vdot\grad)\left(-\frac{\vb*{v}}{T}\right)+v_j\grad\left(-\frac{v_j}{T}\right)\right],\label{S:gradv2}\\
		\vb*{Q}^{\grad\mu}&=\left(\frac{1}{\tau_\mathrm{N}}+\frac{1}{\tau_\mathrm{U}}\right)^{-1}T\frac{u^{(0)}+P}{m^\ast}\grad\left(-\frac{\tilde{\mu}}{T}\right),\label{S:gradmu2}
	\end{align}
\end{subequations}
where we have defined
\begin{equation}
	\langle (\mathcal{N}_4)^2\rangle_\mathrm{eq}\equiv\int[\dd{\vb*{k}}](\hbar\omega_{\vb*{k}})^2N_{\vb*{k}}.
\end{equation}

In summary, the heat current has the following contributions:
\begin{equation}
	\vb*{Q}=\vb*{Q}^{\grad\mu}+\vb*{Q}^{\grad\vb*{v}}+\vb*{Q}^{\grad T}+\vb*{Q}^\tau,
\end{equation}
whose components are given by Eqs.~\eqref{S:heattau} and ~\eqref{S:grad}.

\section{Particle Current}
The same calculations can be performed for the particle currents as
\begin{align}
	\vb*{J}&=\int[\dd{\vb*{k}}]\left(\pdv{\omega_{\vb*{k}}}{\vb*{k}}-\vb*{v}\right)\delta n_{\vb*{k}}\nonumber\\
	&=-\left(\frac{1}{\tau_\mathrm{N}}+\frac{1}{\tau_\mathrm{U}}\right)^{-1}\int[\dd{\vb*{k}}]\left(\pdv{\omega_{\vb*{k}}}{\vb*{k}}-\vb*{v}\right)\left[\pdv{\omega_{\vb*{k}}}{\vb*{k}}\vdot\grad n_{\vb*{k}}^{(0)}+\frac{n_{\vb*{k}}^{(0)}-N_{\vb*{k}}}{\tau_\mathrm{U}}\right]\nonumber\\
	&=\vb*{J}^{\grad T}+\vb*{J}^{\grad\vb*{v}}+\vb*{J}^{\grad\mu}+\vb*{J}^{\tau},
\end{align}
whose components are given by
\begin{subequations}
	\begin{align}
		\vb*{J}^{\tau}&=\frac{1}{\tau_\mathrm{U}}\left(\frac{1}{\tau_\mathrm{N}}+\frac{1}{\tau_\mathrm{U}}\right)^{-1}\int[\dd{\vb*{k}}]\left(\pdv{\omega_{\vb*{k}}}{\vb*{k}}-\vb*{v}\right)N_{\vb*{k}},\\
		\nonumber\\
		\vb*{J}^{\grad T}&=-\left(\frac{1}{\tau_\mathrm{N}}+\frac{1}{\tau_\mathrm{U}}\right)^{-1}T\int[\dd{\vb*{k}}]\left(\pdv{\omega_{\vb*{k}}}{\vb*{k}}-\vb*{v}\right)\hbar\omega_{\vb*{k}}\left\{\pdv{\omega_{\vb*{k}}}{\vb*{k}}\vdot\grad\left(\frac{1}{T}\right)\right\}\pdv{n_{\vb*{k}}^{(0)}}{(\hbar\omega_{\vb*{k}})}\nonumber\\
		&=-\left(\frac{1}{\tau_\mathrm{N}}+\frac{1}{\tau_\mathrm{U}}\right)^{-1}T\int[\dd{\vb*{k}}]\left(\pdv{\omega_{\vb*{k}}}{\vb*{k}}-\vb*{v}\right)\frac{(\hbar\vb*{k}-m^\ast\vb*{v})^2}{2m^\ast}\left\{\left(\pdv{\omega_{\vb*{k}}}{\vb*{k}}-\vb*{v}\right)\vdot\grad\left(\frac{1}{T}\right)\right\}\pdv{n_{\vb*{k}}^{(0)}}{(\hbar\omega_{\vb*{k}})}\nonumber\\
		&\qquad-\left(\frac{1}{\tau_\mathrm{N}}+\frac{1}{\tau_\mathrm{U}}\right)^{-1}T\int[\dd{\vb*{k}}]\left(\pdv{\omega_{\vb*{k}}}{\vb*{k}}-\vb*{v}\right)\frac{(\hbar\vb*{k}-m^\ast\vb*{v})^2}{2m^\ast}\left\{\vb*{v}\vdot\grad\left(\frac{1}{T}\right)\right\}\pdv{n_{\vb*{k}}^{(0)}}{(\hbar\omega_{\vb*{k}})}\nonumber\\
		&\qquad-\left(\frac{1}{\tau_\mathrm{N}}+\frac{1}{\tau_\mathrm{U}}\right)^{-1}T\int[\dd{\vb*{k}}]\left(\pdv{\omega_{\vb*{k}}}{\vb*{k}}-\vb*{v}\right)(\hbar\vb*{k}-m^\ast\vb*{v})\vdot\vb*{v}\left\{\left(\pdv{\omega_{\vb*{k}}}{\vb*{k}}-\vb*{v}\right)\vdot\grad\left(\frac{1}{T}\right)\right\}\pdv{n_{\vb*{k}}^{(0)}}{(\hbar\omega_{\vb*{k}})}\nonumber\\
		&\qquad-\left(\frac{1}{\tau_\mathrm{N}}+\frac{1}{\tau_\mathrm{U}}\right)^{-1}T\int[\dd{\vb*{k}}]\left(\pdv{\omega_{\vb*{k}}}{\vb*{k}}-\vb*{v}\right)(\hbar\vb*{k}-m^\ast\vb*{v})\vdot\vb*{v}\left\{\vb*{v}\vdot\grad\left(\frac{1}{T}\right)\right\}\pdv{n_{\vb*{k}}^{(0)}}{(\hbar\omega_{\vb*{k}})}\nonumber\\
		&\qquad-\left(\frac{1}{\tau_\mathrm{N}}+\frac{1}{\tau_\mathrm{U}}\right)^{-1}T\int[\dd{\vb*{k}}]\left(\pdv{\omega_{\vb*{k}}}{\vb*{k}}-\vb*{v}\right)\frac{m^\ast}{2}\vb*{v}^2\left\{\left(\pdv{\omega_{\vb*{k}}}{\vb*{k}}-\vb*{v}\right)\vdot\grad\left(\frac{1}{T}\right)\right\}\pdv{n_{\vb*{k}}^{(0)}}{(\hbar\omega_{\vb*{k}})}\nonumber\\
		&\qquad-\left(\frac{1}{\tau_\mathrm{N}}+\frac{1}{\tau_\mathrm{U}}\right)^{-1}T\int[\dd{\vb*{k}}]\left(\pdv{\omega_{\vb*{k}}}{\vb*{k}}-\vb*{v}\right)\frac{m^\ast}{2}\vb*{v}^2\left\{\vb*{v}\vdot\grad\left(\frac{1}{T}\right)\right\}\pdv{n_{\vb*{k}}^{(0)}}{(\hbar\omega_{\vb*{k}})},\\
		\nonumber\\
		\vb*{J}^{\grad \vb*{v}}&=-\left(\frac{1}{\tau_\mathrm{N}}+\frac{1}{\tau_\mathrm{U}}\right)^{-1}\int[\dd{\vb*{k}}]\left(\pdv{\omega_{\vb*{k}}}{\vb*{k}}-\vb*{v}\right)\left\{\pdv{\omega_{\vb*{k}}}{\vb*{k}}\vdot\grad\left(-\frac{v_j}{T}\right)\right\}\hbar k_j\pdv{n_{\vb*{k}}^{(0)}}{(\hbar\omega_{\vb*{k}})}\nonumber\\
		&=-\left(\frac{1}{\tau_\mathrm{N}}+\frac{1}{\tau_\mathrm{U}}\right)^{-1}T\int[\dd{\vb*{k}}]\left(\pdv{\omega_{\vb*{k}}}{\vb*{k}}-\vb*{v}\right)\left\{\left(\pdv{\omega_{\vb*{k}}}{\vb*{k}}-\vb*{v}\right)\vdot\grad\left(-\frac{v_j}{T}\right)\right\}(\hbar k_j-m^\ast v_j)\pdv{n_{\vb*{k}}^{(0)}}{(\hbar\omega_{\vb*{k}})}\nonumber\\
		&\qquad-\left(\frac{1}{\tau_\mathrm{N}}+\frac{1}{\tau_\mathrm{U}}\right)^{-1}T\int[\dd{\vb*{k}}]\left(\pdv{\omega_{\vb*{k}}}{\vb*{k}}-\vb*{v}\right)\left\{\vb*{v}\vdot\grad\left(-\frac{v_j}{T}\right)\right\}(\hbar k_j-m^\ast v_j)\pdv{n_{\vb*{k}}^{(0)}}{(\hbar\omega_{\vb*{k}})}\nonumber\\
		&\qquad-\left(\frac{1}{\tau_\mathrm{N}}+\frac{1}{\tau_\mathrm{U}}\right)^{-1}T\int[\dd{\vb*{k}}]\left(\pdv{\omega_{\vb*{k}}}{\vb*{k}}-\vb*{v}\right)\left\{\left(\pdv{\omega_{\vb*{k}}}{\vb*{k}}-\vb*{v}\right)\vdot\grad\left(-\frac{v_j}{T}\right)\right\}m^\ast v_j\pdv{n_{\vb*{k}}^{(0)}}{(\hbar\omega_{\vb*{k}})}\nonumber\\
		&\qquad-\left(\frac{1}{\tau_\mathrm{N}}+\frac{1}{\tau_\mathrm{U}}\right)^{-1}T\int[\dd{\vb*{k}}]\left(\pdv{\omega_{\vb*{k}}}{\vb*{k}}-\vb*{v}\right)\left\{\vb*{v}\vdot\grad\left(-\frac{v_j}{T}\right)\right\}m^\ast v_j\pdv{n_{\vb*{k}}^{(0)}}{(\hbar\omega_{\vb*{k}})},\\
		\nonumber\\
		\vb*{J}^{\grad\mu}&=-\left(\frac{1}{\tau_\mathrm{N}}+\frac{1}{\tau_\mathrm{U}}\right)^{-1}T\int[\dd{\vb*{k}}]\left(\pdv{\omega_{\vb*{k}}}{\vb*{k}}-\vb*{v}\right)\left\{\pdv{\omega_{\vb*{k}}}{\vb*{k}}\vdot\grad\left(-\frac{\tilde{\mu}}{T}\right)\right\}\pdv{n_{\vb*{k}}^{(0)}}{(\hbar\omega_{\vb*{k}})}\nonumber\\
		&=-\left(\frac{1}{\tau_\mathrm{N}}+\frac{1}{\tau_\mathrm{U}}\right)^{-1}T\int[\dd{\vb*{k}}]\left(\pdv{\omega_{\vb*{k}}}{\vb*{k}}-\vb*{v}\right)\left\{\left(\pdv{\omega_{\vb*{k}}}{\vb*{k}}-\vb*{v}\right)\vdot\grad\left(-\frac{\tilde{\mu}}{T}\right)\right\}\pdv{n_{\vb*{k}}^{(0)}}{(\hbar\omega_{\vb*{k}})}\nonumber\\
		&\qquad-\left(\frac{1}{\tau_\mathrm{N}}+\frac{1}{\tau_\mathrm{U}}\right)^{-1}T\int[\dd{\vb*{k}}]\left(\pdv{\omega_{\vb*{k}}}{\vb*{k}}-\vb*{v}\right)\left\{\vb*{v}\vdot\grad\left(-\frac{\tilde{\mu}}{T}\right)\right\}\pdv{n_{\vb*{k}}^{(0)}}{(\hbar\omega_{\vb*{k}})}.
	\end{align}
\end{subequations}
By employing the formula Eqs.~\eqref{S:formula}, these components are expressed in terms of hydrodynamic variables as,
\begin{subequations}
	\begin{align}
		\vb*{J}^{\tau}&=-\frac{1}{\tau_\mathrm{U}}\left(\frac{1}{\tau_\mathrm{N}}+\frac{1}{\tau_\mathrm{U}}\right)^{-1}n ^{(0)}\vb*{v},\\
		\vb*{J}^{\grad T}&=\left(\frac{1}{\tau_\mathrm{N}}+\frac{1}{\tau_\mathrm{U}}\right)^{-1}T\left[\frac{u^{(0)}+P}{m^\ast}+n ^{(0)}\frac{\vb*{v}^2}{2}\right]\grad\left(\frac{1}{T}\right)\nonumber\\
		&\quad+\left(\frac{1}{\tau_\mathrm{N}}+\frac{1}{\tau_\mathrm{U}}\right)^{-1}Tn ^{(0)}\vb*{v}(\vb*{v}\vdot\grad)\left(\frac{1}{T}\right),\\
		\vb*{J}^{\grad\vb*{v}}&=\left(\frac{1}{\tau_\mathrm{N}}+\frac{1}{\tau_\mathrm{U}}\right)^{-1}Tn ^{(0)}\left[(\vb*{v}\vdot\grad)\left(-\frac{\vb*{v}}{T}\right)+v_j\grad\left(-\frac{v_j}{T}\right)\right],\\
		\vb*{J}^{\grad \mu}&=\left(\frac{1}{\tau_\mathrm{N}}+\frac{1}{\tau_\mathrm{U}}\right)^{-1}T\frac{n ^{(0)}}{m^\ast}\grad\left(-\frac{\tilde{\mu}}{T}\right).
	\end{align}
\end{subequations}

\section{transport coefficients}
In order to check the validity of our theory, we investigate the transport coefficients $\mathcal{L}_{ij}$ which are defined as linear response of the particle and heat currents to the gradients of the magnon chemical potential and temperature:
\begin{equation}
	\left(
	\begin{array}{c}
		\vb*{J}   \\
		\vb*{Q}
	\end{array}
	\right)=\left(
	\begin{array}{cc}
		\mathcal{L}_{11} & \mathcal{L}_{12} \\
		\mathcal{L}_{21} & \mathcal{L}_{22}
	\end{array}
	\right)\left(
	\begin{array}{c}
		\grad\left(-\mu/T\right)  \\
		\grad\left(1/T\right) 
	\end{array}
	\right),
\end{equation}

In this section, we assume that the gradients of two ingredients are homogeneous and uniform velocity profile is realized. Therefore, we can ignore the contributions from $\vb*{Q}^{\grad\vb*{v}}$. Furthermore, in the linear response regime, the particle and heat currents are approximated as
\begin{align}
	\vb*{J}&\simeq\left(\frac{1}{\tau_\mathrm{N}}+\frac{1}{\tau_\mathrm{U}}\right)^{-1}T\frac{n ^{(0)}}{m^\ast}\left[\grad\left(-\frac{{\mu}}{T}\right)-\frac{1}{T}\frac{m^\ast\vb*{v}}{\tau_\mathrm{U}}\right],\\
	\vb*{Q}&\simeq\left(\frac{1}{\tau_\mathrm{N}}+\frac{1}{\tau_\mathrm{U}}\right)^{-1}T\frac{u^{(0)}+P}{m^\ast}\left[\grad\left(-\frac{\mu}{T}\right)+\left(\frac{4}{d}+1\right)\frac{\langle(\mathcal{N}_4)^2\rangle_\mathrm{eq}}{u^{(0)}+P}\grad\left(\frac{1}{T}\right)-\frac{1}{T}\frac{m^\ast\vb*{v}}{\tau_\mathrm{U}}\right].
\end{align}
In order to obtain the coefficients $\mathcal{L}_{ij}$, we have to know the drift velocity $\vb*{v}$ by solving the momentum conservation law Eq.~\eqref{S:momentum4}, 
\begin{align}
	\pdv{x_i}P+\pdv{\Pi_{ji}'}{x_j}=-\frac{\rho v_i}{\tau_\mathrm{U}}.\label{S:Euler}
\end{align}
In the following, we neglect $\partial_j\Pi_{ji}'$ because this term only gives higher order contributions. Thus, the magnon fluid is driven by the gradient of the pressure,
\begin{align}
	\grad P&=\left(\frac{d}{2}+1\right)\frac{k_\mathrm{B}T}{\Lambda_{T}^d}\frac{\grad T}{T}\mathrm{Li}_{d/2+1}(z)+\frac{k_\mathrm{B}T}{\Lambda_{T}^d}\grad z\dv{z}\mathrm{Li}_{d/2+1}(z)\nonumber\\
	&=\frac{1}{\Lambda_{T}^d}\left[\left(\frac{d}{2}+1\right)\mathrm{Li}_{d/2+1}(z)k_\mathrm{B}T\frac{\grad T}{T}+\mathrm{Li}_{d/2}(z)\left(\grad\mu-\mu\frac{\grad T}{T}\right)\right]\nonumber\\
	&=\frac{-T}{\Lambda_{T}^d}\left[\left(\frac{d}{2}+1\right)\mathrm{Li}_{d/2+1}(z)k_\mathrm{B}T\grad\left(\frac{1}{T}\right)+\mathrm{Li}_{d/2}(z)\grad\left(-\frac{\mu}{T}\right)\right]\nonumber\\
	&=-T\left[(u^{(0)}+P)\grad\left(\frac{1}{T}\right)+n ^{(0)}\grad\left(-\frac{\mu}{T}\right)\right]\nonumber\\
	&=-Tn ^{(0)}\left[\grad\left(-\frac{\mu}{T}\right)+\alpha\grad\left(\frac{1}{T}\right)\right],
\end{align}
where $\alpha=(u^{(0)}+P)/n ^{(0)}$ and we have used the relation
\begin{equation}
	\dv{z}\mathrm{Li}_{s+1}(z)=\frac{1}{z}\mathrm{Li}_{s}(z).
\end{equation}
Therefore, we obtain the solution for Eq.~\eqref{S:Euler} as
\begin{equation}
	\vb*{v}=\frac{\tau_\mathrm{U}T}{m^\ast}\left[\grad\left(-\frac{\mu}{T}\right)+\alpha\grad\left(\frac{1}{T}\right)\right].
\end{equation}
If the velocity profile becomes nonuniform due to viscous effects, for example the Poiseuille flow, $\vb*{v}$ is of the form
\begin{equation}
	\vb*{v}=\frac{\tau_\mathrm{U}T}{m^\ast}\left[\grad\left(-\frac{\mu}{T}\right)+\alpha\grad\left(\frac{1}{T}\right)\right]\left[1-\frac{\cosh(y/\ell)}{\cosh(w/2\ell)}\right],
\end{equation}
where we consider a finite sample size $w$ in the $y$-direction: $-w/2\leq y\leq w/2$, and $\ell=\sqrt{\nu\tau_\mathrm{U}}$ with the kinematic viscosity $\nu$. For this reason, we introduce a renormalized velocity $\mathcal{V}$ in the velocity profile as
\begin{equation}
	\vb*{v}=\frac{\tau_\mathrm{U}T}{m^\ast}\left[\grad\left(-\frac{\mu}{T}\right)+\alpha\grad\left(\frac{1}{T}\right)\right]\mathcal{V}.
\end{equation}
Here, $\mathcal{V}$ deviates from unity when viscous effects are switched on. By using this quantity, the transport coefficients are obtained as follows:
\begin{subequations}\label{S:coeff}
	\begin{align}
		\mathcal{L}_{11}&=\left(\frac{1}{\tau_\mathrm{N}}+\frac{1}{\tau_\mathrm{U}}\right)^{-1}n ^{(0)}\frac{T}{m^\ast}\left(1-\mathcal{V}\right),\\
		\mathcal{L}_{12}&=\left(\frac{1}{\tau_\mathrm{N}}+\frac{1}{\tau_\mathrm{U}}\right)^{-1}(u^{(0)}+P)\frac{T}{m^\ast}\left(1-\mathcal{V}\right)=\alpha\mathcal{L}_{11},\\
		\mathcal{L}_{21}&=\mathcal{L}_{12},\\
		\mathcal{L}_{22}&=\left(\frac{1}{\tau_\mathrm{N}}+\frac{1}{\tau_\mathrm{U}}\right)^{-1}(u^{(0)}+P)\frac{T}{m^\ast}\left[\left(\frac{4}{d}+1\right)\frac{\langle(\mathcal{N}_4)^2\rangle_\mathrm{eq}}{u^{(0)}+P}-\mathcal{V}\alpha\right],
	\end{align}
\end{subequations}
which obey the Onsager reciprocal relations.

We are now ready to define the spin and heat conductivities as
\begin{equation}
	\hbar(n \vb*{v}+\vb*{J})=\sigma \grad(-\mu),\qquad\vb*{Q}=\kappa \grad(-T).
\end{equation}
In the present case, we set $\mathcal{V}=1$ by neglecting viscous effects, and then obtain two conductivities as
\begin{subequations}\label{S:conduct}
	\begin{align}
		\sigma &=\hbar\left[\frac{n \tau_\mathrm{U}}{m^\ast}+\frac{1}{T}\mathcal{L}_{11}\right]=\frac{n ^{(0)}\hbar\tau_\mathrm{U}}{m^\ast},\\
		\kappa &=\frac{1}{T^2}\left[\mathcal{L}_{22}-\frac{\mathcal{L}_{12}\mathcal{L}_{21}}{\mathcal{L}_{11}}\right]=\frac{1}{T^2}(\mathcal{L}_{22}-\alpha^2\mathcal{L}_{11})\nonumber\\
		&=\left(\frac{1}{\tau_\mathrm{N}}+\frac{1}{\tau_\mathrm{U}}\right)^{-1}\frac{1}{T^2}\frac{T}{m^\ast}\left[\left(\frac{4}{d}+1\right)\langle(\mathcal{N}_4)^2\rangle_\mathrm{eq}-\alpha^2n ^{(0)}\right].
	\end{align}
\end{subequations}
Substituting analytic form of $n ^{(0)}$ and $\langle(\mathcal{N}_4)^2\rangle_\mathrm{eq}$ into Eqs.~\eqref{S:conduct}, we finally obtain the magnonic Lorentz number in the hydrodynamic regime:
\begin{align}
	\left.\frac{\kappa }{\sigma T }\right|_\mathrm{hydro}&=\left(\frac{1}{\tau_\mathrm{N}}+\frac{1}{\tau_\mathrm{U}}\right)^{-1}\frac{1}{\tau_\mathrm{U}}\frac{k_\mathrm{B}^2}{\hbar}\left[\frac{(d+4)(d+2)}{4}\frac{\mathrm{Li}_{d/2+2}(z)}{\mathrm{Li}_{d/2}(z)}-\left(\frac{d+2}{2}\right)^2\frac{\mathrm{Li}_{d/2+1}^2(z)}{\mathrm{Li}_{d/2}^2(z)}\right]\nonumber\\
	&=\frac{\tau_\mathrm{N}}{\tau_\mathrm{N}+\tau_\mathrm{U}}\left.\frac{\kappa }{\sigma T }\right|_\mathrm{non-int}.
\end{align}

\section{detailed calculations}
Under the assumption that magnons have a parabolic dispersion: $\hbar\omega_{\vb*{k}}=\hbar^2\vb*{k}^2/2m^\ast$, the density of states $D(\epsilon)$ is proportional to $\epsilon^{d/2-1}$ for $d$-dimensional systems. We list the detailed analytical calculations for hydrodynamic variables in the following,
\begin{equation}
	n ^{(0)}=\int[\dd{\vb*{k}}]n_{\vb*{k}}^{0}(\vb*{r},t)=D(1)\int\dd\epsilon\frac{\epsilon^{{d}/{2}-1}}{e^{(\epsilon-\mu)/k_\mathrm{B}T}-1}=D(1)(k_\mathrm{B}T)^{{d}/{2}}\int\dd{x}\frac{x^{d/2-1}}{e^x/z-1}=\frac{1}{\Lambda_{T}^d}\mathrm{Li}_{d/2}(z),
\end{equation}
where $z=e^{\mu/k_\mathrm{B}T}$ is the fugacity of magnons and $\Lambda_T=\sqrt{\frac{2\pi\hbar^2}{m^\ast k_\mathrm{B}T}}$ is the thermal de Broglie wavelength. Furthermore, we have used the formula
\begin{align}
	\int\dd{x}\frac{x^{d/2-1}}{e^x/z-1}&=\int\dd{x}ze^{-x}\frac{x^{d/2-1}}{1-ze^{-x}}=\int\dd{x}\sum_{n=1}^\infty(ze^{-x})^nx^{d/2-1}=\sum_{n=1}^\infty\frac{z^n}{n^{d/2}}\int\dd{\tau}e^{-\tau}\tau^{d/2-1}\nonumber\\
	&=\Gamma\left(\frac{d}{2}\right)\mathrm{Li}_{d/2}(z),
\end{align}
where $\mathrm{Li}_s(z)=\sum_{n=1}^\infty z^n/n^s$ and $\Gamma(s)$ are the polylogarithm and gamma functions. Performing the same procedures, we obtain
\begin{gather}
	P\equiv-k_\mathrm{B}T\int[\dd{\vb*{k}}]\log[1-e^{-\beta(\hbar\omega_{\vb*{k}}-\mu)}]=\frac{k_\mathrm{B}T}{\Lambda_{T}^d}\mathrm{Li}_{d/2+1}(z)=\frac{2}{d}u^{(0)},\\
	u^{(0)}=\int[\dd{\vb*{k}}]\hbar\omega_{\vb*{k}}n_{\vb*{k}}^0=D(1)(k_\mathrm{B}T)^{d/2+1}\mathrm{Li}_{d/2+1}(z)\Gamma\left(\frac{d}{2}+1\right)=\frac{d}{2}\frac{k_\mathrm{B}T}{\Lambda_{T}^{d/2}}\mathrm{Li}_{d/2+1}(z),\\
	\langle (\mathcal{N}_4)^2\rangle_\mathrm{eq}\equiv\int[\dd{\vb*{k}}](\hbar\omega_{\vb*{k}})^2N_{\vb*{k}}=\left(\frac{d}{2}+1\right)\frac{d}{2}\frac{(k_\mathrm{B}T)^2}{\Lambda_{T}^d}\mathrm{Li}_{d/2+2}(z).    
\end{gather}

\end{document}